\documentclass[reprint,aps,graphicx,pre,showpacs]{revtex4-2}
\usepackage{amsmath,amssymb,amsfonts}     

\usepackage{graphicx,bm}
  \usepackage{paralist}
  \usepackage{epstopdf}
  \usepackage{graphics} 
 \usepackage[colorlinks=true]{hyperref}

\renewcommand{\theequation}{\arabic{section}.\arabic{equation}}

\usepackage{float}
\usepackage{epsfig}
\usepackage{esint}
\usepackage{chemfig}
\usepackage{bm,braket}
\usepackage{soul}
\usepackage{hyperref}

\renewcommand{\theequation}{\arabic{section}.\arabic{equation}}

\newcommand{\calP}{{\mathcal P}}

\newcommand{\pt}{\widetilde{\rho}}
\newcommand{\R}{{\mathbb R}}

\newcommand{\e}{{\mathrm e}}
\newcommand{\E}{{\mathbb E}}

\newcommand{\calT}{{\mathcal T}}
\newcommand{\calF}{{\mathcal F}}

\newcommand{\p}{\widetilde{p}}

\newcommand{\f}{\widetilde{f}}
\newcommand{\q}{\widetilde{q}}

\newcommand{\wrho}{\widetilde{\rho}}
\newcommand{\wphi}{\widetilde{\phi}}

\renewcommand{\S}{\widetilde{S}}

\newcommand{\dd}{\,\textup{d}}

\newcommand{\fT}{{\mathfrak T}}

\begin{document}

 \title{Renewal theory for a run-and-tumble particle with stochastic resetting and a sticky boundary}

\author{Paul C. Bressloff and Samantha Linn}
\address{Department of Mathematics, Imperial College London, London SW7 2AZ, UK}

\begin{abstract} 
We consider a run-and-tumble particle (RTP) with stochastic resetting confined to the half line $[0,\infty)$ with a sticky boundary at $x=0$. In the bulk the RTP tumbles at a constant rate $\alpha>0$ between velocity states $\pm v$ with $v>0$ and randomly resets to its initial position and orientation $(x_0,k_0)\in(\mathbb{R}^+,\pm)$. When the RTP reaches the target at $x=0$ it attaches to the boundary for a random waiting time before either detaching and continuing to navigate the bulk domain or (permanently) entering the target. These events are the analogs of adsorption, desorption, and absorption of a particle by a partially reactive surface in physical chemistry. We use renewal theory to characterize the particle trajectory in terms of successive binding events under two distinct desorption protocols: via resetting to $(x_0,k_0)$ and via continuous movement from $x=0$ with velocity $+v$. First we derive the nonequilibrum stationary state (NESS) in the case of no absorption and characterize the accumulation at the boundary. Second, we compute the mean first passage time (MFPT) statistics. In addition to observing the usual unimodal dependence of the MFPT on bulk resetting, both the NESS and MFPT strongly depend on the initial orientation $k_0$ and the desorption protocol. For instance, if the initial orientation is toward the boundary, we find that the desorption-induced resetting protocol can reduce the MFPT more effectively than the non-resetting desorption protocol. We also show how matching the desorption kinetics with the bulk resetting or tumbling rate introduces a trade-off between minimizing the adsorption and absorption times. In this setting we find that the desorption protocol which minimizes the absorption MFPT for a given set of parameters is almost always the opposite of that favored when desorption and bulk kinetics are not the same.
\end{abstract}

\maketitle
\section{Introduction}

One of the characteristic features of ensembles of self-propelled particles (active matter) is that the particles tend to accumulate at a confining wall even in the absence of attractive particle interactions \cite{Ramaswamy10,Vicsek12,Roman12,Bechinger16}. A simple example of such a phenomenon is a run-and-tumble particle (RTP) confined to a one-dimensional (1D) interval. Away from the boundary walls, the particle randomly switches at a Poisson rate $\alpha$ between a left-moving and a right-moving constant velocity state. Whenever the particle collides with a wall (adsorption) it remains there until it reverses its velocity state and re-enters the bulk domain (desorption). Thus each wall acts as a so-called sticky boundary. It is also possible that whilst attached to a wall the RTP is permanently removed from the system (absorption) before detaching. The first passage time (FPT) problem for absorption of an RTP at a sticky boundary has been analysed extensively elsewhere \cite{Angelani17,Angelani23}, and has been generalized to include the effects of diffusion \cite{Malakar18,Bressloff25} and encounter-based models of non-Markovian absorption \cite{Bressloff23,Bressloff25d}. There have also been several studies of FPT problems for RTPs with stochastic resetting \cite{Evans18,Bressloff20,Santra20a,Bressloff25c} or a confining potential \cite{Dhar19}. However, none of these include the effects of sticky boundaries.

In this paper we develop a renewal formulation of a stochastically resetting RTP on the half-line $[0,\infty)$ with a sticky boundary at $x=0$ by extending recent work on diffusion-mediated adsorption/desorption/absorption \cite{Bressloff25a,Bressloff25b}. (The latter itself builds upon a renewal formulation of reversible adsorption in physical chemistry \cite{Grebenkov23,Scher23}.) We proceed by decomposing sample paths into an alternating sequence of bulk diffusion with resetting and first-return to $x=0$, followed by adsorption/desorption at the wall. This sequence continues until there is an absorption event. Using the fact that each first-return to $x=0$ acts as a renewal of the stochastic process, we construct renewal equations that relate the probability density and FPT density for absorption to the corresponding quantities for irreversible adsorption. The renewal equations take the form of implicit integral equations that can be solved explicitly using Laplace transforms and the convolution theorem. A crucial step in the renewal formulation is ensuring that the particle restarts in a state that avoids immediate re-adsorption. In the case of a Brownian particle, this has been implemented by taking the sticky boundary to be partially adsorbing \cite{Grebenkov23,Scher23,Bressloff25a} or by resetting the particle to its initial position away from the wall following desorption \cite{Bressloff25b}.

One particular feature of an RTP is that the particle desorbs from the wall in a right-moving state that naturally avoids immediate re-adsorption. (This would no longer be true for an RTP subject to diffusion as in Refs. \cite{Malakar18,Bressloff25}.) However, as we show in the paper, the analysis of the corresponding renewal equations is considerably more involved than for desorption-induced resetting, say, since we have to condition on the time $T$ of the first velocity reversal or bulk resetting event following desorption, whose splitting probabilities are determined by their respective rates, and then averaging with respect to $T$. A major goal of the paper is to compare the effects of desorption-induced resetting versus desorption directly from the boundary on an RTP with bulk resetting. In particular, we compare the respective nonequilibrium stationary states (NESS) in the case of a non-absorbing sticky boundary and the mean FPTs (MFPTs) in the case of an absorbing sticky boundary. In the latter case, we show how desorption-induced resetting can reduce the MFPT when the particle resets to a left-moving state.

 The structure of the paper is as follows. In Sect. II we consider the simpler problem of an RTP on the half-line with a non-absorbing sticky boundary at $x=0$ and no resetting. We begin by considering the standard forward Kolmogorov partial differential equation (PDE) for the pair of probability densities associated with the left-moving and right-moving state, respectively. In order to develop the renewal formulation, we solve the Laplace transformed Kolmogorov equation under the initial condition that the particle leaves the wall in the right-moving state. We then reinterpret the sticky boundary as a reactive surface with reversible adsorption, under the assumption that, following each desorption event, the RTP restarts its motion at $x=0$ in the right-moving  state. We construct the first renewal equation relating the probability densities for the left-moving and right-moving states in the presence of a sticky boundary to the corresponding densities for irreversible adsorption. The latter are calculated by averaging with respect to the time $T$ of the first tumble event following desorption. We highlight one of the major advantages of the renewal formulation, namely, that we can include a non-exponential waiting time density for the duration of an adsorbed or bound state prior to desorption/absorption. We then establish the equivalence of the renewal equation and the Kolmogorov equation in the case of an exponential waiting time density. The latter recovers the standard sticky boundary with constant rates of desorption and absorption.

In Sect. III we introduce stochastic resetting in the bulk while maintaining a non-absorbing sticky boundary at $x=0$. That is, in the bulk domain the particle spontaneously resets to its initial position and velocity state at a Poisson rate $r$. Assuming that the RTP restarts from the wall in the right-moving state, we construct the corresponding first renewal equation and calculate the densities for irreversible adsorption by averaging with respect to the time $T$ of the first tumble or bulk resetting event following desorption. We then use the renewal equation to calculate the NESS and show how the NESS at a position $x$ can be interpreted in terms of the expected fraction of time an RTP spends in a neighborhood of $x$ between adsorption events. We explore the dependence of the NESS and the probability of being adsorbed on various model parameters including the bulk resetting rate, the tumbling rate, the initial position and the initial velocity state. Finally, we compare our results to the much simpler example of desorption-induced resetting. In particular, we show how the latter can result in greater accumulation at the sticky boundary when the RTP resets to a left-moving state. 
 
In Sect. IV we consider a resetting RTP on the half-line with a partially absorbing sticky boundary at $x=0$. Continuing to allow desorption thereby introduces splitting probabilities for whether the bound particle absorbs or desorbs after some random waiting time at the boundary. We proceed by constructing the FPT density for absorption as a first renewal integral equation with respect to desorption. Again we consider both desorption from the boundary and desorption-induced resetting. We thus determine the possible interplay between bulk and desorption-induced resetting in optimizing the MFPT. In particular, we find that while the non-resetting desorption protocol is relatively more effective at enhancing the NESS near the boundary and maximizing the time fraction bound to the wall, the desorption-induced resetting protocol can yield lower absorption MFPTs when the RTP resets to the left-moving state and the tumbling rate is relatively small. If instead we equate the desorption rate to the bulk tumbling or resetting rate, depending on the desorption protocol, we discover that the parameter regime in which desorption-induced resetting yields lower absorption MFPTs is effectively the complement of that from before. These results and other subsidiary implications are made rigorous and thoroughly described throughout the rest of the paper.

One final comment is in order. Throughout the main text we simply write down the appropriate renewal equations based on heuristic arguments, and give probabilistic interpretations of the various terms. This is also the approach taken in related papers on Brownian particles with sticky boundaries \cite{Grebenkov23,Scher23,Bressloff25a,Bressloff25b}. In Appendix A we give a detailed first principles derivation of the renewal equations for an RTP based on renewal theory. The basic idea is to express the propagator or FPT density as an expectation with respect to an empirical measure, which is then conditioned on the first arrival time at the boundary. Such a formulation could be adapted to other examples of continuous stochastic processes with renewals.

\section{RTP on the half-line with a (non-absorbing) sticky boundary}

\begin{figure}[t!]
\centering
\includegraphics[width=8cm]{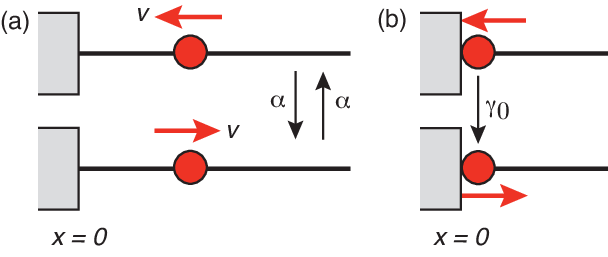}
\caption{An RTP confined to the half-line $[0,\infty)$ by a non-absorbing sticky wall at $x=0$. (a) In the bulk domain the particle randomly switches between a left-moving and a right-moving constant velocity state at a Poisson rate $\alpha$. (b) Whenever the RTP collides with the wall in the left-moving state it binds to the wall (adsorbs). It subsequently unbinds (desorbs) at a constant rate $\gamma_0$ and reenters the bulk domain in the right-moving state.}
\label{fig1}
\end{figure}

Consider an RTP confined to the half-line by a non-absorbing sticky wall as illustrated in Fig. \ref{fig1}. Away from the wall, the particle randomly switches at a constant rate $\alpha$ between two constant velocity states $\sigma(t)\in \{+,-\}$ with corresponding velocities $\sigma(t)v$, $v>0$, see Fig. \ref{fig1}(a). Let $\rho_{k|k_0}(x,t|x_0)$ be the probability density that at time $t$ particle is at $X(t)=x>0$ and in velocity state $\sigma(t)=k$ given that $X(0)=x_0$ and $\sigma(0)=k_0$. The associated forward Kolmogorov equation is
\begin{subequations}
\begin{align}
\label{RTPa}
\frac{\partial \rho_{k|k_0}}{\partial t}&=-vk \frac{\partial \rho_{k|k_0}}{\partial x}-\alpha \rho_{k|k_0}+\alpha \rho_{-k|k_0}.
\end{align}
Suppose that whenever the RTP hits the boundary at $x=0$ it sticks to the wall and remains in the bound state until desorbing at a rate $\gamma_0$, see Fig. \ref{fig1}(b).
Let $q(t)$ denote the probability that at time $t$ the particle is in the bound state. The boundary condition at $x=0$ takes the form
\begin{align}
v\rho_{+|k_0}(0,t|x_0)=\gamma_0q(t),
\label{RTPc}
\end{align}
with $q(t)$ evolving according to the equation
\begin{equation}
\label{RTPd}
\frac{dq}{dt}=v \rho_{-|k_0}(0,t|x_0)- \gamma_0 q(t).
\end{equation}
(Note that $q(t)$ depends on $x_0$ and $k_0$, and we assume $q(0)=0$.)
\end{subequations}

 \subsection{Solution in Laplace space}
Suppose that the RTP starts at $x_0>0$ in the left-moving state. That is, $k_0=-$ and
\begin{equation}
\rho_{+|-}(x,0|x_0)=0,\quad \rho_{-|-}(x,0|x_0)=\delta(x-x_0).
\end{equation}
(The more typical initial condition is $k_0=\pm$ with equal probability. However, in anticipation of later developments, we take $k_0=-$.) 
Laplace transforming Eqs.\ \eqref{RTPa}--\eqref{RTPd} then gives
\begin{subequations}
\begin{align} \label{iLTPa}
  v\frac{\partial \pt_{+|-} }{\partial x}+(\alpha +s)\pt_{+|-} 
  -\alpha \pt_{-|-}  &=0,\\
\label{iLTPb}
 -v\frac{\partial \pt_{-|-} }{\partial x}+(\alpha +s)\pt_{-|-} -\alpha \pt_{+|-}  &=\delta(x-x_0)
\end{align}
with the boundary conditions 
\begin{align}
v \pt_{+|-}(0,s|x_0)&=\gamma_0 \q(s), \\
v \pt_{-|-}(0,s|x_0)&=(s+\gamma_0) \q(s).
\end{align}
\end{subequations}
It follows that
\begin{equation}
\label{GH}
\pt_{-|-}(0,s|x_0)=\frac{s+\gamma_0}{\gamma_0}\pt_{+|-}(0,s|x_0).
\end{equation}
Multiplying both sides of Eq.\ \eqref{iLTPa} by the operator $-v\partial_x+(\alpha+s)$ and using Eq.\ \eqref{iLTPb} yields the second-order equation for $\pt_{+|-}(x,s|x_0)$,
 \begin{align} \label{SOE}
 &D(s)\frac{\partial^2 \pt_{+|-} }{\partial x^2}-s\pt_{+|-}
   =-\frac{\alpha}{s+2\alpha } \delta(x-x_0),
\end{align}
with the effective diffusivity $D(s)=v^2/(s+2\alpha)$. Moreover, setting $x=0$ in Eq.\ \eqref{iLTPa} and using Eq.\ \eqref{GH} yields the boundary condition for the right-moving density
\begin{equation} \label{bc_right}
D(s) \frac{\partial \pt_{+|-}}{\partial x} \Big|_{x=0} =\gamma(s) \pt_{+|-} ,
 \end{equation}
 with
 \begin{equation}
 \gamma(s)=\frac{v(\alpha -\gamma_0)s}{(s+2\alpha)\gamma_0} .
 \end{equation}
It follows from Eqs.\ \eqref{SOE} and \eqref{bc_right} that
\begin{align}
\label{pt1}
& \pt_{+|-}(x,s|x_0)= \frac{\alpha}{s+2\alpha } \frac{1}{2\sqrt{sD(s)}}\\
 &\quad \times \bigg [\e^{-\sqrt{s/D(s)}|x-x_0|}+\theta(s)\e^{-\sqrt{s/D(s)}(x+x_0)}\bigg ],\nonumber
 \end{align}
 where
 \begin{equation}
 \theta(s)=\frac{\sqrt{sD(s)}-\gamma(s)}{\sqrt{sD(s)}+\gamma(s)}.\end{equation} 
The corresponding solution for the left-moving density $\pt_{-|-}(x,s|x_0)$ can then be obtained from Eq.\ \eqref{iLTPa} for $x>0$ and Eq.\ \eqref{GH} when $x=0$.

\subsection{Starting at the wall} 
The above analysis has to be modified if the particle unbinds from the wall at time $t=0$, that is, $x_0=0$ and $k_0=+$. Let $T$ denote the exponentially distributed time for the first tumble. For a given $T$, we have the relation
\begin{align}
&\rho_{k|+}^{(T)}(x,t|0)  \\
&=\delta_{k,+}\delta(x-vt)\Theta(T-t)+\rho_{k|-} (x,t-T|vT)\Theta(t-T) \nonumber,
\end{align}
where $\Theta(t)$ is the Heaviside function. That is, the RTP travels a distance $vt$ in the right-moving state for $ t \leq T$ and then switches to the left-moving state at time $T$. Averaging with respect to $T$ then shows that
\begin{align}
\rho_{k|+}(x,t|0) &=\alpha \int_0^{\infty} \e^{-\alpha T} \bigg [\delta_{k,+}\delta(x-vt)\Theta(T-t)\nonumber  \\
&\qquad +\rho_{k|-}(x,t-T|vT)\Theta(t-T)\bigg ] \dd T \nonumber\\
 &=\delta_{k,+}\delta(x-vt)\e^{-\alpha x/v} \nonumber \\
 &\quad +\alpha \int_0^t \e^{-\alpha T} \rho_{k|-}(x,t-T|vT) \dd T . 
\end{align}
 Laplace transforming both sides, we have
\begin{align}
\wrho_{k|+}(x,s|0)
 &=v^{-1}\delta_{k,+}\e^{-(s+\alpha) x/v} +\widetilde{\calP}_k(x,s),
 \end{align}
 where
 \begin{align}
\widetilde{\calP}_k(x,s)&\equiv \alpha \int_0^{\infty}\dd t\, \e^{-st}\int_0^t \dd T\, \e^{-\alpha T} \rho_{k|-}(x,t-T|vT)   \nonumber \\
&=\alpha \int_0^{\infty}\dd T\, \e^{-\alpha T} \int_T^{\infty} \dd t\, \e^{-st}\rho_{k|-}(x,t-T|vT)  \nonumber \\
&=\alpha \int_0^{\infty}\dd T\, \e^{-(s+\alpha) T} \wrho_{k|-}(x,s|vT).
\end{align}
The final step is to substitute for $ \wrho_{k|-}(x,s|vT)$ using the solutions obtained from Sect. II.A. As a simple illustration, set $k=-$ and $x=0$. We then have $\wrho_{-|+}(0,s|0)
=\widetilde{\calP}_-(0,s)$ with
\begin{align} \label{res1}
&\widetilde{\calP}_-(0,s) \\
&=\alpha \int_0^{\infty}\dd T\, \e^{-(s+\alpha) T} \wrho_{-|-}(0,s|vT) \nonumber \\
&=\alpha \frac{s+\gamma_0}{\gamma_0}\int_0^{\infty}\dd T\, \e^{-(s+\alpha) T} \wrho_{+|-}(0,s|vT)\nonumber \\
&=\frac{s+\gamma_0}{\gamma_0} \bigg [ \frac{\alpha^2}{s+2\alpha }\bigg] \frac{1+\theta(s)}{2\sqrt{sD(s)}}  \int_0^{\infty}\dd T\, \e^{-(s+\alpha+\sqrt{s/D(s)}v) T} \nonumber \\
&=\frac{s+\gamma_0}{\gamma_0} \bigg [ \frac{\alpha^2}{s+2\alpha }\bigg ]  \frac{1}{s+\alpha+\sqrt{s/D(s)}v} \frac{1}{\sqrt{sD(s)}+\gamma(s)}.\nonumber
\end{align}
 
 \subsection{Renewal equation}
 We now rewrite the forward Kolmogorov equations \eqref{RTPa}--\eqref{RTPd} in the form of a renewal equation along analogous lines to our recent formulation of Brownian motion in the presence of partially reactive boundaries \cite{Bressloff25a,Bressloff25b}. One major advantage of the renewal equation is that it allows us to incorporate a more general model of desorption. In particular, we will assume that when the particle is adsorbed, it remains bound for a random time $\tau$ generated from a waiting time density $\phi(\tau)$. In the case of the exponential density $\phi(\tau)=\gamma_0\e^{-\gamma_0\tau}$, we recover the case of a constant desorption rate $\gamma_0$. Applications of the renewal equation will be explored in subsequent sections.
  
Let $p_{k|k_0}(x,t|x_0)$ denote the probability densities in the case of irreversible adsorption ($\gamma_0=0$), which evolve according to the forward Kolmogorov equation
 \begin{subequations}
\begin{align}
\label{iRTPa}
\frac{\partial p_{k|k_0}}{\partial t}&=-vk \frac{\partial p_{k|k_0}}{\partial x}-\alpha p_{k|k_0}+\alpha p_{-k|k_0}
\end{align}
subject to the totally adsorbing boundary condition
\begin{align}
p_{+|k_0}(0,t|x_0)  =0.
\label{iRTPc}
\end{align}
\end{subequations}
The survival probability for the adsorption time with respect to irreversible adsorption is
\begin{equation}
S_{k_0}(x_0,t)=\int_0^{\infty}p_{k_0}(x,t|x_0)\dd x,
\label{Sp0}
\end{equation}
where $p_{k_0}=p_{+|k_0}+p_{-|k_0}$, and the corresponding FPT density is 
\begin{equation}
f_{k_0}(x_0,t)\equiv-\frac{\partial S_{k_0}(x_0,t)}{\partial t}=v p_{-|k_0}(0,t|x_0).
\end{equation}
The renewal equations relate the densities $\rho_{k|k_0}(x,t|x_0)$ in the presence of desorption (reversible adsorption) to the corresponding quantities $p_{k|k_0}(x,t|x_0)$ and $f_{k_0}(x_0,t)$ without desorption (irreversible adsorption). 

After each desorption event, the particle restarts its motion at $x=0$ in the right-moving state as analyzed in Sect. II.B. The corresponding renewal equation for the propagator takes the form
\begin{align}
\label{ren1}
  &\rho_{k|k_0}(x,t|x_0)=p_{k|k_0}(x,t|x_0)\\
  & +\int_0^t\dd \tau' \int_{\tau'}^t \dd\tau\,  \rho_{k|+}(x,t-\tau|0)  \phi(\tau-\tau') f_{k_0}(x_0,\tau').\nonumber
 \end{align}
The first term on the right-hand side of the renewal equation \eqref{ren1} for the density represents the contribution from all sample paths that start at $x_0$ in velocity state $k_0$ and have not been adsorbed over the interval $[0,t]$. On the other hand, the second term represents all sample paths that are first adsorbed at a time $\tau' $ with probability $f_{k_0}(x_0,\tau')d\tau'$, remain in the bound state until desorbing at time $\tau$ with probability $ \phi(\tau-\tau')d\tau$, after which the particle may bind an arbitrary number of times before reaching $x$ at time $t$. One can view the renewal equation as sewing together successive rounds of adsorption and desorption, as illustrated in Fig. \ref{fig2}. (Note that Eq.\ \eqref{ren1} is a special case of the general renewal Eq. \eqref{srenp} derived in Appendix A with $\pi_d=1$ (no absorption), $(x^*,k^*)=(0,k_+)$ (no resetting after desorption) and no bulk resetting.)
 
 \begin{figure}[t]
\centering
\includegraphics[width=8cm]{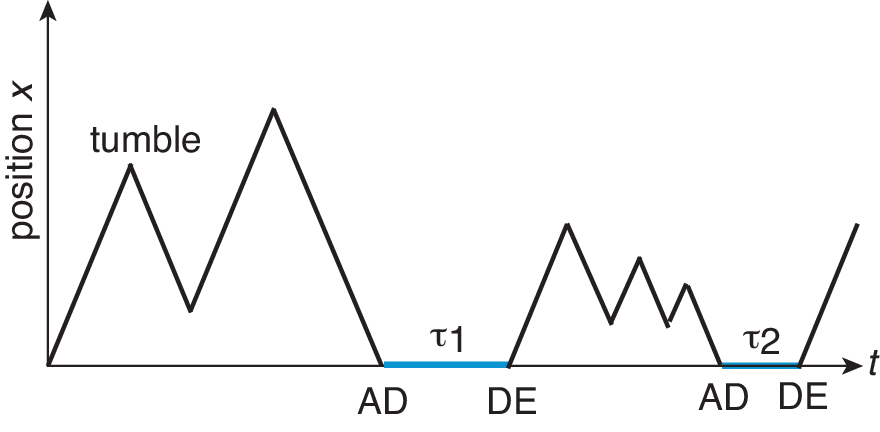}
\caption{Example trajectory of an RTP with a (non-absorbing) sticky wall at $x=0$. Each time the particle collides with the wall in the left-moving state it is adsorbed (AD). After the $j$th collision, the particle remains attached or bound to the wall for a random waiting time $\tau_j$, after which it is desorbed (DE) and re-enters the bulk domain in the right-moving state.}
\label{fig2}
\end{figure}

 Laplace transforming Eq.\ \eqref{ren1} gives
\begin{align}
\wrho_{k|k_0}(x,s|x_0)&=\p_{k|k_0}(x,s|x_0) \\
&\quad + \wrho_{k|+}(x,s|0)  \wphi(s) \f_{k_0}(x_0,s).\nonumber
 \end{align}
 Setting $(x_0,k_0)=(0,+)$ and rearranging shows that
 \begin{equation}
 \label{rel1}
 \wrho_{k|+}(x,s|0)=\frac{\p_{k|+}(x,s|0)}{1-  \wphi(s) \f_+(0,s)}
 \end{equation}
 and, hence,
 \begin{align}
 \label{LTren1}
\wrho_{k|k_0}(x,s|x_0)&=\p_{k|k_0}(x,s|x_0) \\
&\quad +\frac{ \wphi(s) \f_{k_0}(x_0,s)}{1- \wphi(s) \f_+(0,s)}\p_{k|+}(x,s|0).\nonumber
 \end{align}
We see that the contribution from paths that desorb at least once involves the quantities $\p_{k|+}(x,s|0)$ and $\f_+(0,s)=v\p_{-|+}(0,s|0)$ associated with irreversible adsorption. The Laplace transforms $\p_{k|+}(x,s|0)$ can be obtained directly from the solutions for $\wrho_{k|+}(x,s|0)$ by taking the limit $\gamma_0\rightarrow 0$. For example, Eq.\ \eqref{pt1} yields
\begin{align} \label{ppm}
& \p_{+|-}(x,s|x_0)= \frac{\alpha}{s+2\alpha } \frac{1}{2\sqrt{sD(s)}}\\
 &\quad \times \bigg [\e^{-\sqrt{s/D(s)}|x-x_0|}-\e^{-\sqrt{s/D(s)}(x+x_0)}\bigg ],\nonumber
 \end{align}
 for $x_0>0$, which then determines $ \p_{-|-}(x,s|x_0)$ via the analog of Eq.\ \eqref{iLTPa}. Similarly, taking $\gamma_0\rightarrow 0$ in Eq.\ \eqref{res1} implies that
\begin{align}
\p_{-|+}(0,s|0)
&=  \frac{\alpha \sqrt{sD(s)}/v}{(s+\alpha)\sqrt{sD(s)}+sv}  .
\end{align}

 To check the equivalence of the renewal equation \eqref{ren1} and the corresponding forward Kolmogorov equations \eqref{RTPa}--\eqref{RTPd}, we compare their solutions in Laplace space for $x=0$, $k=-$ and a constant rate of desorption $\gamma_0$ for which $\wphi(s)=\gamma_0/(s+\gamma_0)$. First, setting $x=0$ and $k=-$ in Eq.\ \eqref{rel1} gives
 \begin{align} \label{wrhomp}
 \wrho_{-|+}(0,s|0)&=\frac{\p_{-|+}(0,s|0)}{1 -\wphi(s)v \p_{-|+}(0,s|0)}
 \end{align}
 We have also used $\f_+(0,s)=v\p_{-|+}(0,s|0)$. Second, substituting for $\p_{-|+}(0,s|0)$ and $\wphi(s)$ yields
   \begin{align}
   \label{res2}
 &\wrho_{-|+}(0,s|0)\\
 &=\frac{\alpha \sqrt{sD(s)}(s+\gamma_0)/v}{[\sqrt{sD(s)}(\alpha+s)+vs](s+\gamma_0)-\alpha\gamma_0 \sqrt{sD(s)}}.\nonumber 
 \end{align}
Finally, after some algebra, it can be shown that Eq.\ \eqref{res2} is equivalent to Eq.\ \eqref{res1}. 

Notice that the contribution of the waiting time before desorption in Eq.\ \eqref{wrhomp} is entirely contained in the multiplicative factor $\wphi(s)$. Accommodating other finite-moment waiting time distributions therefore merely requires a single-term substitution. Suppose for instance that the waiting time density is gamma distributed, i.e.
\begin{align} \label{gamdist}
	\phi(\tau) = \frac{\gamma_0(\gamma_0 \tau)^{\mu-1}e^{-\gamma_0 \tau}}{\Gamma(\mu)}
\end{align}
with $\mu>0$ and $\Gamma(\mu)$ denoting the gamma function. The Laplace transform of Eq.\ \eqref{gamdist} is $\wphi(s)=(\gamma_0/(s+\gamma_0))^\mu$, hence substitution into Eq.\ \eqref{wrhomp} yields
\begin{align}
	&\wrho_{-|+}(0,s|0)\\
 	&= \frac{\alpha \sqrt{sD(s)}/v}{\sqrt{sD(s)}(\alpha+s)+vs - \alpha (\gamma_0/(s+\gamma_0))^\mu \sqrt{sD(s)}}. \nonumber
\end{align}
Comparable substitutions yield the particle densities for any $(k,k_0)\in(\pm,\pm)$.

 \setcounter{equation}{0}
 \section{RTP with stochastic resetting: NESS for a non-absorbing sticky wall}
 
  \begin{figure}[t!]
\centering
\includegraphics[width=8cm]{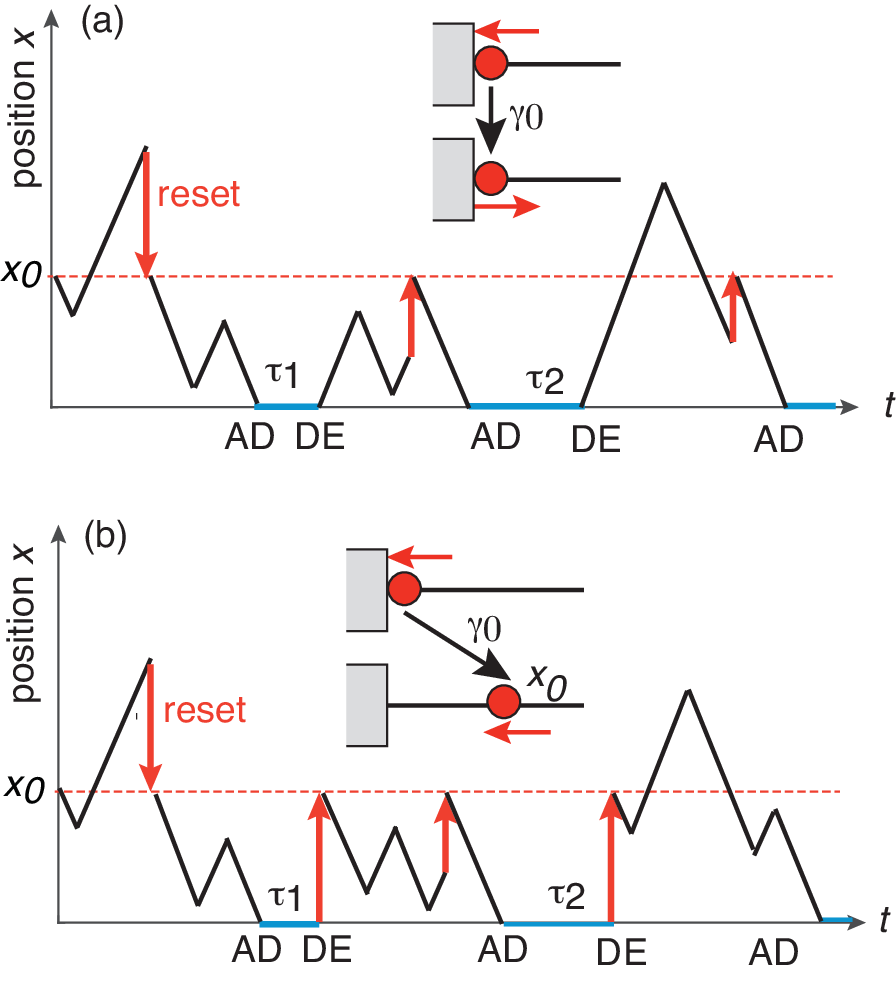} 
\caption{Example trajectories of an RTP on the half-line with a (non-absorbing) sticky wall at $x=0$ and instantaneous stochastic resetting at a rate $r$ to $(x_0,k_0)$. For the sake of illustration we take $k_0=-$. The sequence of waiting times in the bound state are given by $\{\tau_1,\tau_2,\ldots \}$. (a) The particle continues from $x=0$ in the right-moving state after each desorption event. (b) The particle resets to $x_0$ in the velocity state $k_0=-$ immediately after each desorption event. }
\label{fig3}
\end{figure}

 As our first application of the renewal formulation, we calculate the non-equilibrium stationary state (NESS) for an RTP on the half-line with a non-absorbing sticky wall at $x=0$ and stochastic resetting. That is, we assume that when the RTP is in the bulk domain $(0,\infty)$,
its position $X(t)$ is reset to its initial location $x_0$ at random times generated from a Poisson distribution with constant rate $r\geq 0$. We also assume that the discrete state $\sigma(t)$ is reset to the initial velocity state $k_0$. (This differs from the randomized velocity reset rule originally considered in Ref. \cite{Evans18}.) Following our recent analysis of Brownian motion \cite{Bressloff25b}, we distinguish between two different protocols after each desorption event. The first assumes that the particle continues from $x=0$ in the $+$ state as in Sect. II, whereas the second assumes that the particle resets to its initial state $(x_0,k_0)$. In the latter case, there are two distinct types of resetting events: spontaneous resetting in the bulk domain and resetting from the boundary following desorption. The two desorption protocols are illustrated in Fig. \ref{fig3}. 

\subsection{No reset after desorption}
First suppose that the particle continues from $x=0$ in the $+$ state after desorption. Since the particle desorbs in a different state to its bulk resetting state, $(x_0,k_0)$, we introduce the following notation: given a state $(y_0,j_0)$ we denote by $\rho_{r,k|j_0}(x,t|y_0)$ and $p_{r,k|j_0}(x,t|y_0)$ the corresponding probability densities for reversible and irreversible adsorption, respectively, starting from $(y_0,j_0)$. We also use this notation to describe first passage times, $\tau_{r,j_0}(y_0)$ and ${\mathcal T}_{r,j_0} (y_0)$. Throughout, the subscript $r$ indicates bulk resetting to $(x_0,k_0)$ at a rate $r$. If $(y_0,j_0)=(x_0,k_0)$ then the initial and bulk resetting states are the same.

For $y_0>0$ the probability density $p_{r,k|j_0}(x,t|y_0)$ satisfies the forward Kolmogorov equation
\begin{subequations}
\label{CKHH2}
\begin{eqnarray}
\frac{\partial p_{r,k|j_0}}{\partial t}&=&-vk \frac{\partial p_{r,k|j_0}}{\partial x}-(\alpha +r) p_{r,k|j_0}+\alpha  p_{r,-k|j_0}
\nonumber \\
&&\quad + rS_{r,j_0}(y_0,t) \delta(x-x_0)\delta_{k,k_0}, \\
 0&=&{p}_{r,+|j_0}(0,t|y_0),
\end{eqnarray}
\end{subequations}
where $S_{r,j_0}(y_0,t)$ is the survival probability for the adsorption time,
\begin{equation}
 S_{r,j_0}(y_0,t)=\int_0^{\infty}p_{r,j_0}(x,t|y_0)\dd x
 \end{equation}
and $p_{r,j_0}\!=\!p_{r,+|j_0}+p_{r,-|j_0}$. If $\tau_{r,j_0}(y_0)=\inf\{t\!>\!0,\ X(t)\!=\!0\}$ denotes the FPT for irreversible adsorption, then 
\begin{equation}
 {\rm Prob}[\tau_{r,j_0}(y_0)\leq t] =1-S_{r,j_0}(y_0,t).
 \end{equation} 
  The FPT density is thus given by
\begin{equation}
\label{surv}
 f_{r,j_0}(y_0,t)=-\frac{\partial S_{r,j_0}(y_0,t)}{\partial t}= vp_{r,-|j_0}(0,t|y_0).
\end{equation}
The renewal equation for the corresponding particle which starts at $(x_0,k_0)$ thereby takes the same form as Eq.\ \eqref{ren1} except that $p\rightarrow p_r$ etc.:
\begin{align}
\label{renr2}
  &\rho_{r,k|k_0}(x,t|x_0)=p_{r,k|k_0}(x,t|x_0)\\
  & +\int_0^t\dd\tau' \int_{\tau'}^t \dd\tau\,  \rho_{r,k|+}(x,t-\tau|0)  \phi(\tau-\tau') f_{r,k_0}(x_0,\tau').\nonumber
 \end{align}

We will calculate the NESS $\rho_{r,k|k_0}^*$ by working in Laplace space and using the relation
\begin{equation} \label{rela}
\rho_{r,k|k_0}^*(x|x_0)\equiv \lim_{t\rightarrow \infty}\rho_{r,k|k_0}(x,t|x_0)=\lim_{s\rightarrow 0}s\wrho_{r,k|k_0}(x,s|x_0) . 
\end{equation}
The NESS depends on $(x_0,k_0)$ since this is also the reset point. Laplace transforming the renewal Eq.\ \eqref{renr2} and rearranging gives
 \begin{align}
 \label{rLTren1}
\wrho_{r,k|k_0}(x,s|x_0)&=\p_{r,k|k_0}(x,s|x_0) \\
&\quad +\frac{ \wphi(s) \f_{r,k_0}(x_0,s)}{1- \wphi(s) \f_{r,+}(0,s)}\p_{r,k|+}(x,s|0).\nonumber
\end{align}
Since
 \begin{equation}
\lim_{s\rightarrow 0} \wphi(s)=1, \quad \lim_{s\rightarrow 0} \f_{r,j_0}(y_0,s)=1
\end{equation}
for all $y_0>0$ and for $(y_0,j_0)=(0,+)$,
it follows that
\begin{align}
 \label{rLTrs02}
\rho_{r,k|k_0}^*(x|x_0)
&=\lim_{s\rightarrow 0}\frac{ s\p_{r,k|+}(x,s|0)}{1- \wphi(s) \f_{r,+}(0,s)} \nonumber\\
&= - \frac{\p_{r,k|+}(x,0|0)}{\displaystyle \left . \frac{d}{ds}\bigg (\wphi(s) \f_{r,+}(0,s)\bigg )\right |_{s=0}}\nonumber \\
&=\frac{\p_{r,k|+}(x,0|0)}{\langle \tau\rangle +\calT_{r,+}(0)}
\end{align}
with 
\begin{align}
\langle \tau\rangle =\int_0^{\infty} \tau \phi(\tau)\dd\tau =-\left . \frac{d}{ds} \wphi(s) \right |_{s=0}
\end{align}
and
\begin{align}
{\mathcal T}_{r,j_0} (y_0) =-\left . \frac{d}{ds} \f_{r,j_0}(y_0,s) \right |_{s=0} = \S_{r,j_0}(y_0,0).
\label{Tr0}
\end{align}
In other words, $\langle \tau \rangle$ is the mean waiting time for desorption and ${\mathcal T}_{r,j_0}(y_0)$ is the MFPT for irreversible adsorption under bulk resetting to $(x_0,k_0)$ with initial state $(y_0,j_0)$.

The last line of Eq. (\ref{rLTrs02}) has the following interpretation. Since
\begin{align}
	\p_{r,k|+}(x,0|0)=\int_0^{\infty}p_{r,k|+}(x,t|0)\,\textup{d}t,
\end{align}
we see that $\p_{r,k|+}(x,0|0)\textup{d}x$ is the expected amount of time the particle spends in the infinitesimal interval $[x,x+\textup{d}x]$ averaged with respect to all sample paths that start at $x=0$ in the right-moving state and are adsorbed after a mean time $\calT_{r,+} (0)$. If $\langle \tau \rangle =0$, then the boundary is totally reflecting rather than sticky. This implies that we can identify $\p_{r,k|+}(x,0|0)/\calT_{r,+}(0)$ as the NESS for an RTP with bulk resetting and a totally reflecting wall at $x=0$. Hence, Eq. (\ref{rLTrs02}) can be rewritten in the form
 \begin{align}
 \label{rLTrs022}
\rho_{r,k|k_0}^*(x|x_0)
&=\frac{\calT_{r,+}(0)}{\langle \tau\rangle +{\mathcal T}_{r,+}(0)}\rho_{r,k|+}^*(x|0)_{\rm reflecting}.
\end{align}
The multiplicative factor on the right-hand side is the mean fraction of time the RTP spends in the bulk rather than stuck to the boundary.

\subsection{Starting from the boundary}

We see from Eq. (\ref{rLTrs02}) that calculating the NESS reduces to solving for $\p_{r,k|+}(x,s|0)_{s=0}$ and the associated MFPT $T_{r,+}(0)$, which requires extending the analysis of Sect. II.B. Let $T$ denote the exponentially distributed time for the first tumbling or reset event, hence $T$ has probability density $\psi(T)=(\alpha+r)\e^{-(\alpha+r)T} $. For a given $T$, we have the relation
\begin{widetext}
\begin{align}
&p_{r,k|+}^{(T)}(x,t|0)   
 =\delta_{k,+}\delta(x-vt)\Theta(T-t)+\bigg [\frac{\alpha}{\alpha +r}p_{r,k|-} (x,t-T|vT) 
 +\frac{r}{\alpha +r}p_{r,k|k_0} (x,t-T|x_0)\bigg ]\Theta(t-T)  .
\end{align}
That is, the RTP travels a distance $vT$ in the right-moving state and then either switches with probability $\alpha/(\alpha+r)$ to the left-moving state or resets with probability $r/(\alpha+r)$ to $(x_0,k_0)$. Averaging with respect to $T$ then yields
\begin{align}
p_{r,k|+}(x,t|0) &= \int_0^{\infty} \e^{-(\alpha +r)T} \bigg [(\alpha+r)\delta_{k,+}\delta(x-vt)\Theta(T-t) +\bigg (\alpha p_{r,k|-}(x,t-T|vT)+r p_{r,k|k_0}(x,t-T|x_0)\bigg )\Theta(t-T)\bigg ]\dd T \nonumber\\
 &=\delta_{k,+}\delta(x-vt)\e^{-(\alpha+r) x/v} +\int_0^t \e^{-(\alpha+r) T}\bigg (\alpha  p_{r,k|-}(x,t-T|vT) +r p_{r,k|k_0}(x,t-T|x_0)\bigg )\dd T .
\end{align}
 Laplace transforming both sides, we have
\begin{align} \label{pp00}
\p_{r,k|+}(x,s|0)
 &=v^{-1}\delta_{k,+}\e^{-(r+s+\alpha) x/v} +  \int_0^{\infty}\dd T\, \e^{-(r+s+\alpha) T}\bigg (\alpha \p_{r,k|-}(x,s|vT)+r\p_{r,k|k_0}(x,s|x_0)\bigg ).
 \end{align}
We determine an expression for $\p_{r,k|-}(x,s|vT)$ in Eq.\ \eqref{pp00} via a first renewal equation for $p_{r,k|-}(x,t|vT)$,
\begin{align} \label{renew-vT}
	&p_{r,k|-}(x,t|vT) = e^{-rt} p_{k|-}(x,t|vT)  
	 + r\int_0^t \dd\tau e^{-r\tau} S_{-}(vT,\tau) p_{r,k|k_0}(x,t-\tau|x_0). 
\end{align}
The first term on the right-hand side of Eq.\ \eqref{renew-vT} represents the contribution from all sample paths that start at position $vT$ moving left and have not reset during the interval $[0,t]$. The second term represents all sample paths that are first reset at a time $\tau$ and thereafter reset an arbitrary number of times before reaching $x$ at time $t$. Laplace transforming Eq.\ \eqref{renew-vT} gives
\begin{align} \label{p2}
	\p_{r,k|-}(x,s|vT)  = \p_{k|-}(x,s+r|vT)+r\S_-(vT,s+r) \p_{r,k|k_0}(x,s|x_0) .
\end{align}
Substituting back into Eq.\ \eqref{pp00} and setting $s=0$, we have
\begin{align} \label{pp0}
\p_{r,k|+}(x,0|0) =v^{-1}\delta_{k,+}\e^{-(r+\alpha) x/v} + \int_0^{\infty}\dd T\, \e^{-(r+\alpha) T}\bigg [\alpha  \p_{k|-}(x,r|vT)+\bigg (1+\alpha \S_{-}(vT,r)\bigg ) r\p_{r,k|k_0}(x,0|x_0) \bigg ].
 \end{align}
\end{widetext}

 Finally, integrating Eq.\ \eqref{pp0} with respect to $x\in [0,\infty)$, summing over $k$, and using Eq.\ \eqref{Tr0} shows that
\begin{align} \label{Trp0}
\calT_{r,+}(0)&=\frac{1+r\calT_{r,k_0}(x_0) + \alpha\calT_{r,-}(v/(r+\alpha))}{r+\alpha}.
\end{align}
The expression for $\calT_{r,+}(0)$ in Eq.\ \eqref{Trp0} has a clear physical interpretation: The average time before the first resetting or tumbling event is given by the inverse sum of their rates. This pivotal event then happens on average at position $x=v/(r+\alpha)$ with splitting probabilities $r/(r+\alpha)$ and $\alpha/(r+\alpha)$, respectively.

Expressions for $\S_{-}(vT,r)$ and $\p_{k|-}(x,r|vT)$ can be obtained from the analysis of Sect.\ IIB; see Eqs.\ \eqref{Sp0} and \eqref{ppm}. What remains is determining $\p_{r,k|k_0}(x,0|x_0)$ and $\calT_{r,k_0}(x_0)$ by solving the forward Kolmogorov Eq.\ \eqref{CKHH2} in Laplace space. This was previously carried out in Ref.\ \cite{Bressloff25c}. For completeness we provide a simplified version of the calculation in Appendix B. We thus obtain the following expressions. First taking the limit $s\!\rightarrow \!0$ in Eq.\ \eqref{pLT} gives
\begin{align}
 \label{pLTk0m}
\p_{r,-k_0|k_0}(x,0|x_0)&=\frac{1}{v}\bigg ( [1+r \calT_{r,k_0}(x_0)  ]G(x,r|x_0)\nonumber \\
&\quad \qquad +\delta_{k_0,+}\e^{-\Lambda(r)x} \bigg )
\end{align}
where $\Lambda(r)=v^{-1}\sqrt{r(2\alpha+r)}$
 and
\begin{equation}
G(x,r|x_0)=\frac{\alpha}{2v\Lambda(r)}\big [\e^{-\Lambda(r)|x-x_0|} -\e^{-\Lambda(r)(x+x_0)}\big ].
\end{equation}
Similarly taking the limit $s\!\rightarrow\! 0$ in Eq.\ \eqref{ord1} yields
 \begin{align}
  \label{pLTk0p}
&\p_{r,k_0|k_0}(x,0|x_0) \\
&= \frac{v}{\alpha} \bigg [-k_0\frac{\partial \p_{r,-k_0|k_0}(x,0|x_0) }{\partial x}+\frac{\alpha+r}{v}\p_{r,-k_0|k_0}(x,0|x_0) \bigg ] .\nonumber 
  \end{align}
  Finally we have from Eq.\ \eqref{Tplus} that for $k_0=+$
  \begin{align}
\calT_{r,k_0} (x_0) =\frac{1}{r}\! \left (\frac{\sqrt{r(2\alpha+r)}+\alpha+r}{\alpha} \e^{\Lambda(r)x_0}-1\right ) \label{Tpma}
\end{align}
and from Eq.\ \eqref{Tminus} for $k_0=-$
\begin{align}
	\calT_{r,k_0}(x_0)&=\frac{1}{r} \left [ \e^{\Lambda(r)x_0}-1\right ]. \label{Tpmb}
\end{align}

We can now complete the calculation of the NESS \eqref{rLTrs02}. First, substituting Eqs.\ \eqref{pLTk0m} and \eqref{pLTk0p} into Eq.\ \eqref{pp0}, we analytically evaluate the integral with respect to $T$. (The resulting expression is rather involved and not further illuminating so we do not include it.) We thus obtain the term $\p_{r,k|+}(x,0|0)$ appearing in the numerator of Eq. \eqref{rLTrs02}. Second, substituting Eqs.\ \eqref{Tpma} and \eqref{Tpmb} into Eq.\ \eqref{Trp0} determines the term $\calT_{r,+}(0)$ appearing in the denominator of Eq.\ \eqref{rLTrs02}. Finally, as in Sect.\ II(c), the waiting time distribution contributes to the particle density only up to a single substitution. If the waiting times are rate $\gamma_0$ exponentially distributed, then $\langle\tau\rangle$ in Eq.\ \eqref{rLTrs02} is simply $1/\gamma_0$. Alternatively if these times are gamma distributed with rate $\gamma_0>0$ and shape $\mu>0$, then $\langle\tau\rangle=\mu/\gamma_0$. Regardless of this choice, the NESS depends merely on the mean of this waiting time.

\subsection{Reset after desorption}
The analysis is considerably simplified if the particle immediately resets after desorption as in Fig.\ \ref{fig3}(a). We continue to identify the bulk resetting state with the initial condition $(x_0,k_0)$. The renewal equation for $\rho_{r,k|k_0}(x,t|x_0)$ thereby takes the form
\begin{align}
\label{renr1}
  &\rho_{r,k|k_0}(x,t|x_0)=p_{r,k|k_0}(x,t|x_0)\\
  & +\int_0^t\dd\tau' \int_{\tau'}^t \dd\tau\,  \rho_{r,k|k_0}(x,t-\tau|x_0)  \phi(\tau-\tau') f_{r,k_0}(x_0,\tau').\nonumber
 \end{align}
 This follows from Eq.\ \eqref{srenp} with $\pi_d=1$ and $(x^*,k^*)=(x_0,k_0)$. We again calculate $\rho_{r,k|k_0}^*$ by working in Laplace space. Laplace transforming Eq.\ \eqref{renr1} and rearranging gives
\begin{align}
\wrho_{r,k|k_0}(x,s|x_0)
&=\frac{ \p_{r,k|k_0}(x,s|x_0)}{1- \wphi(s) \f_{r,k_0}(x_0,s)} .
\end{align}
It follows from Eq.\ \eqref{rela} that
\begin{align}
 \label{rLTrs0}
\rho_{r,k|k_0}^*(x|x_0) =\frac{\p_{r,k|k_0}(x,0|x_0)}{\langle \tau\rangle +\calT_{r,k_0}(x_0)}.
\end{align}
Substitution of Eqs. (\ref{pLTk0m}) and (\ref{pLTk0p}) yields the NESS,
\begin{subequations}
\begin{align}
 \label{NESSmp}
&\rho_{r,-k_0|k_0}^*(x|x_0)
= \frac{1}{v(\langle \tau\rangle +\calT_{r,k_0}(x_0))}\\
&\quad \times \Big[(1+r\calT_{r,k_0}(x_0))G(x,r|x_0)+\delta_{k_0,+}\e^{-\Lambda(r)x}\Big],\nonumber \\
 \label{NESSpp}
&\rho_{r,k_0|k_0}^*(x|x_0)
=  \frac{v}{\alpha} \bigg [ -k_0\frac{\partial \rho^*_{r,-k_0 |k_0}(x|x_0) }{\partial x}\\
	&\qquad+\lambda(r)\rho^*_{r,-k_0|k_0}(x|x_0) \bigg ].\nonumber
\end{align}
\end{subequations}

\subsection{Results}

\begin{figure*}[t!]
\centering
\includegraphics[width=14.93cm]{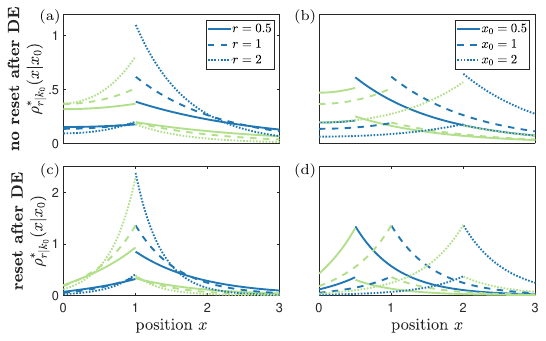}
\caption{NESS of the bulk resetting RTP for (a,b) no resetting after desorption as considered in Sect.\ IIIA,B and (c,d) desorption-induced resetting to $(x_0,k_0)$ as considered  in Sect.\ IIIC. In the first column $x_0=1$ and we vary the bulk resetting rate; in the second column $r=1$ and we vary the initial and resetting position. Other parameters are $\alpha=1$, $\langle \tau\rangle =1$, and $v=2$. Dark blue curves indicate $k_0=+$ and light green curves indicate $k_0=-$.}
\label{fig:NESS}
\end{figure*}

We illustrate the NESS for both desorption protocols in Fig.\ \ref{fig:NESS}. In both cases we observe a striking difference in the NESS with $k_0=+$ compared to $k_0=-$ owing to the flux discontinuity at $x_0$ which endows the NESS with a significant jump favoring one side of the domain from $x_0$. This difference is accentuated by more frequent resetting (larger $r$) since resetting events restore the RTP with its initial orientation. The presence of a boundary induces an asymmetry in $k_0=\pm$ that grows inversely with $x_0$. These observations thus far extend to both desorption protocols nearly identically.

Such qualitative similarities between desorption protocols suggest that practical differences between desorption protocols are subtle. The most notable difference in the NESS is the mass concentrated near the boundary. Clearly the non-resetting protocol gives rise to relatively more mass near the boundary for both $k_0=\pm$ since desorption is modeled by continuous movement from $x=0$. In anticipation of the following section, which considers the sticky boundary to be partially absorbing, one might expect that this enhanced mass near the boundary always has the effect of reducing absorption MFPTs. It turns out that this is only sometimes true.

\begin{figure*}[t!]
\centering
\includegraphics[width=14.93cm]{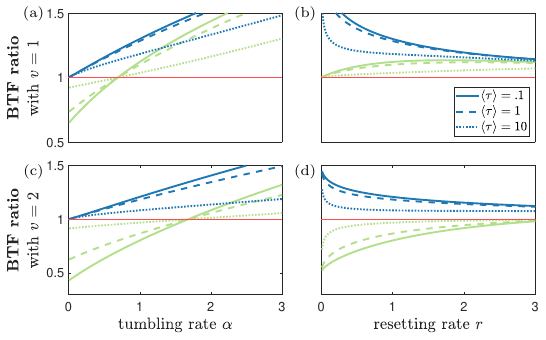}
\caption{Ratio of the bound time fraction (BTF) of the non-resetting desorption protocol to the desorption-induced resetting protocol for various mean waiting times $\langle\tau\rangle$ with (a,b) $v=1$ and (c,d) $v=2$. Data below unity (horizontal red line) implies that the desorption-induced resetting protocol yields a relatively higher BTF. In the first column $r=1$ and we vary the tumbling rate; in the second column $\alpha=1$ and we vary the bulk resetting rate. Other parameters are $x_0=1$ and $v=2$. Dark blue curves indicate $k_0=+$ and light green curves indicate $k_0=-$.}
\label{fig:BTF}
\end{figure*}

A hint toward the breakdown of this conjecture is revealed in Fig.\ \ref{fig:BTF} where we show how the bound time fraction (BTF), which is the average fraction of time spent stuck to the boundary rather than in the bulk, is affected by the bulk resetting and tumbling rates. Unsurprisingly, the non-resetting desorption protocol always yields a higher BTF when the initial orientation is right-moving ($k_0=+$). However, when $k_0=-$, we see from Figs.\ \ref{fig:BTF}(a,c) that desorption-induced resetting yields a higher BTF so long as the tumbling rate is relatively low compared to $v/x_0$. We can explain this by the fact that, upon desorption, the reset RTP moves toward the boundary and is relatively unlikely to tumble while the non-reset RTP moves away from the boundary and is also unlikely to tumble. This phenomenon appears to hold true for all bulk resetting rates $r$, as inferred from the relationship between the $k_0\!=\!-$ curves in Figs.\ \ref{fig:BTF}(b,d). As we will see in Sect.\ IV the values $(\alpha,r)$ for which the BTF ratio in Fig.\ \ref{fig:BTF} dips below one is closely related to those for which desorption-induced resetting yields faster absorption MFPTs.

\clearpage
\pagebreak

\setcounter{equation}{0}
 \section{RTP with stochastic resetting: FPT for absorption at a sticky wall}

 \begin{figure}[t!]
\centering
\includegraphics[width=6cm]{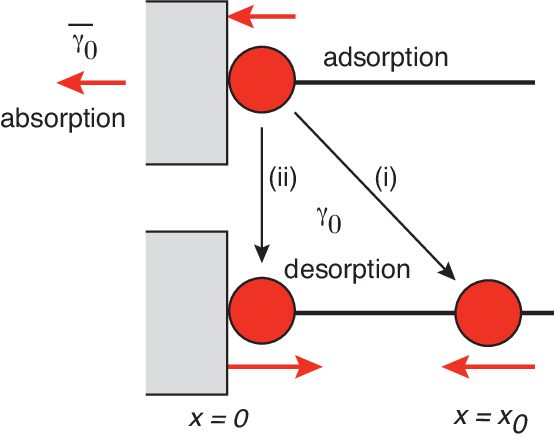}
\caption{Reactions at a partially absorbing sticky wall. Whenever the RTP collides with the wall in the left-moving state it binds to the wall (adsorbs). It then subsequently either unbinds (desorbs) at a constant rate $\gamma_0$ or is permanently removed from the system at a rate $\overline{\gamma}_0$ (absorbs). After desorption it either (i) resets to $x_0$ in the velocity state $k_0=-$, say, or (ii) continues from the wall in the right-moving state. }
\label{fig5}
\end{figure}

As a second application of renewal theory, we analyze the FPT problem for an RTP with resetting on the half-line with a partially absorbing sticky wall at $x=0$, which is illustrated in Fig. \ref{fig5}. As in Sect. III, we will assume that when the particle is adsorbed, it remains bound for a random time $\tau$ generated from a waiting time density $\phi(\tau)$. However, now the particle either desorbs with a splitting probability $\pi_d$ or is permanently absorbed with probability $\pi_b=1-\pi_d$. In the exponential case, we have
 \begin{equation}
 \label{ephi}
 \phi(\tau)=\gamma\e^{-\gamma \tau},\quad \gamma =\gamma_0+\overline{\gamma}_0,
 \end{equation}
 with associated splitting probabilities
 \begin{equation}
 \label{split}
 \pi_d=\frac{\gamma_0}{\gamma_0+\overline{\gamma}_0},\quad \pi_b=\frac{\overline{\gamma}_0}{\gamma_0+\overline{\gamma}_0}.
 \end{equation}
 Similar to our recent renewal formulation of Brownian motion with sticky boundaries \cite{Bressloff25a,Bressloff25b}, it is necessary to modify the renewal equation for the propagator and to supplement it with a renewal equation relating the FPT density for absorption, $\calF_{r,k_0}(x_0,t)$, with the FPT density for adsorption, $f_{r,k_0}(x_0,t)$. The general renewal equations, which are derived in Appendix A, are given by Eqs.\ \eqref{srenp} and \eqref{srenf}, respectively. In Sect.\ IV we focus on the renewal equation for the FPT density.

\subsection{No reset after desorption}

First suppose that the particle continues from $x=0$ in the $+$ state after desorption, see scenario (ii) in Fig. \ref{fig5}.Setting $(x^*,k^*)=(0,k_+)$ in Eq. (\ref{srenf}), the renewal equation for the FPT density is
 \begin{align}
  &\calF_{r,k_0}(x_0,t)=\pi_b\int_0^t\dd \tau  {\phi}(t-\tau) f_{r,k_0}(x_0,\tau)    \label{ren200}\\
&+\pi_d \int_0^t\dd\tau' \int_{\tau'}^t \dd\tau\,  {\calF}_{r,+}(0,t-\tau) \phi(\tau-\tau') f_{r,k_0}(x_0,\tau'). \nonumber
 \end{align}
The first term on the right-hand side represents all sample paths that are first adsorbed at time $\tau$ and are subsequently absorbed at time $t$ without desorbing, which occurs with probability $\pi_b{\phi}(t-\tau)f_{r,k_0}(x_0,\tau)\dd\tau \dd t$. In a complementary fashion, the second term sums over all sample paths that are first adsorbed at time $\tau'$, desorb at time $\tau$ and are absorbed at time $t$ following an arbitrary number of additional adsorption events.

The Laplace transform of Eq.\ \eqref{ren200} is given by
\begin{align}
	\widetilde{\calF}_{r,k_0}(x_0,s) &= \pi_b\wphi(s)\f_{r,k_0}(x_0,s) \label{wcalFn}\\
	&\qquad + \pi_d\wphi(s)\f_{r,k_0}(x_0,s)\widetilde{\calF}_{r,+}(0,s).\nonumber
\end{align}
We ultimately use Eq.\ \eqref{wcalFn} to determine the absorption time moments. To do so we first need to formulate an expression for $\widetilde{\calF}_{r,+}(0,s)$, which can be achieved with its own renewal equation,

\begin{align}
	  &\calF_{r,+}(0,t) = \pi_b\int_0^t\dd \tau  {\phi}(t-\tau) f_{r,+}(0,\tau)    \label{wcalFp}\\
	&+\pi_d \int_0^t\dd\tau' \int_{\tau'}^t \dd\tau\,  {\calF}_{r,+}(0,t-\tau) \phi(\tau-\tau') f_{r,+}(0,\tau'). \nonumber
\end{align}
Taking the Laplace transform of Eq.\ \eqref{wcalFp} and solving for $\widetilde{\calF}_{r,+}(0,s)$ yields
\begin{align}
	\widetilde{\calF}_{r,+}(0,s) &= \frac{\pi_b\wphi(s)\f_{r,+}(0,s)}{1 - \pi_d\wphi(s)\f_{r,+}(0,s)}, \label{wcalFnp}
\end{align}
which we substitute into Eq.\ \eqref{wcalFn} in order to determine the FPT moments,
\begin{align}
	\fT_{r,k_0}^{(n)}(x_0) &:= \int_0^{\infty} t^n \calF_{r,k_0}(x_0,t)\dd t\nonumber \\
	&= \Big( \!-\!\frac{d}{ds} \Big)^n \widetilde{\calF}_{r,k_0}(x_0,s) \Big|_{s=0}.
\end{align}
Assuming moreover that $\phi$ is not heavy-tailed, we substitute the series expansions
\begin{align}
	\f_{r,+}(0,s) &\sim 1 - s \calT_{r,+}(0) + \frac{s^2}{2}\calT^{(2)}_{r,+}(0) + \mathcal{O}(s^3)\nonumber \\
	\f_{r,k_0}(x_0,s) &\sim 1 - s \calT_{r,k_0}(x_0) + \frac{s^2}{2}\calT^{(2)}_{r,k_0}(x_0) + \mathcal{O}(s^3)\nonumber \\
	\wphi(s) &\sim 1 - s\langle\tau\rangle + \frac{s^2}{2}\langle\tau^2\rangle + \mathcal{O}(s^3) \label{srs}
\end{align}
into Eq.\ \eqref{wcalFnp} and Taylor expanding in powers of $s$ to find
\begin{align}
	\label{mfptnrad}
	\fT^{(1)}_{r,k_0}(x_0) = \calT_{r,k_0}(x_0) + \langle\tau\rangle + \Gamma\big(\calT_{r,+}(0) + \langle\tau\rangle\big),
\end{align}
where $\calT_{r,+}(0)$ is given by Eq.\ \eqref{Trp0}, $\calT_{r,k_0}(x_0)$ is given by Eqs.\ \eqref{Tpma} and \eqref{Tpmb} for $k_0=\pm$,  and $\Gamma\equiv\pi_d/\pi_b$.

The physical interpretation of Eq.\ \eqref{mfptnrad} is immediate; the first term on the right-hand side is the mean adsorption time from $x_0$, the second term is the mean absorption time following the final adsorption event, and the third term is the mean time accrued due to (intermediate) desorption events. In particular, the probability of exactly $n$ desorption events is $p_n:=\pi_b\pi_d^n$, hence its mean is precisely
\begin{align}
	\overline{n} := \sum_{n=0}^{\infty} np_n =  \sum_{n=0}^{\infty} n\pi_b\pi_d^n  = \Gamma.
\end{align}
Moreover, the mean time between excursions is exactly $\calT_{r,+}(0) + \langle\tau\rangle$. 

\begin{figure*}[t!]
\centering
\includegraphics[width=17.8cm]{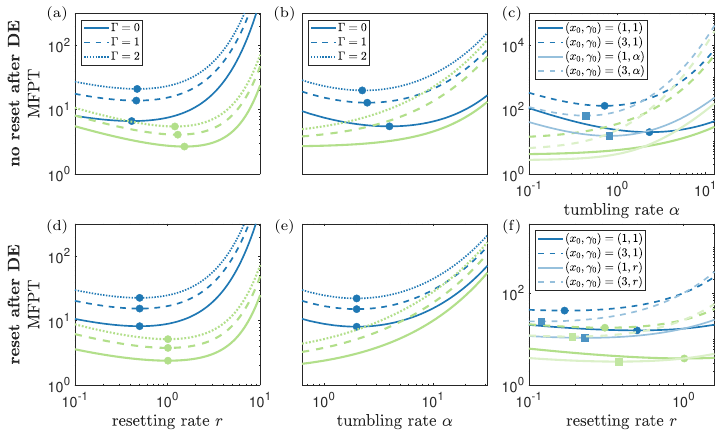}
\caption{Absorption MFPT for (a-c) no resetting after desorption as considered in Sect.\ IVA,B and (d-f) desorption-induced resetting to $(x_0,k_0)$ as considered in Sect.\ IVC. In the first column $\alpha=1$ and we vary the resetting rate; in the second column $r=1$ and we vary the tumbling rate. The third column illustrates how the MFPTs evolve when (c) in the non-resetting desorption protocol, the desorption rate and bulk tumbling rate are the same and when (f) in the desorption-induced resetting protocol, the desorption rate and bulk resetting rate are the same. Blue curves indicate $k_0=+$ and green curves indicate $k_0=-$. Minima are labeled with markers. Other parameters are $x_0=1$, $v=2$, and $\overline{\gamma}_0=1$.}
\label{fig:MFPT}
\end{figure*}

Higher-order moments have a more nuanced dependence on system parameters. Collecting the second-order terms in the Taylor expansion of Eq.\ \eqref{wcalFnp} yields the second-order FPT moment
\begin{align}
	\label{mfpt2nrad}
	&\fT^{(2)}_{r,k_0}(x_0) = (1+\Gamma)\big(\calT^{(2)}_{r,k_0}(x_0) + \langle\tau^2 \rangle +2\langle\tau\rangle \calT_{r,k_0}(x_0)\big)\nonumber\\
	&\hspace{.2em} +\Gamma\big( \calT^{(2)}_{r,+}(0) - \calT^{(2)}_{r,k_0}(x_0) + 2 \calT_{r,+}(0) \big[\calT_{r,k_0}(x_0)- \calT_{r,+}(0)\big] \big)\nonumber \\
 	&\hspace{.2em} + 2\Gamma(1+\Gamma)\big( \calT_{r,+}(0)+\langle\tau\rangle \big)^2
\end{align}
with
\begin{align}
	\calT^{(2)}_{r,k_0}(x_0) = \frac{d^2 \f_{r,k_0}(x_0,s)}{ds^2}\Big|_{s=0} = v\frac{d^2 \p_{r,-|k_0}(0,s|x_0) }{ds^2} \Big|_{s=0}
\end{align}
and similarly for $\calT^{(2)}_{r,+}(0)$. Note from Eqs.\ \eqref{mfptnrad} and \eqref{mfpt2nrad} that in the absence of desorption ($\Gamma=0$) the FPT variance is simply the sum of the variances from irreversible adsorption and from waiting at the boundary,
\begin{align}
	\fT^{(2)}_{r,k_0}&(x_0) - (\fT^{(1)}_{r,k_0}(x_0))^2 \label{vargam0}\\
 	&= \calT^{(2)}_{r,k_0}(x_0) - (\calT_{r,k_0}(x_0))^2 + \langle\tau^2 \rangle - \langle\tau\rangle^2.\nonumber
\end{align}
We illustrate these results in Fig.\ \ref{fig:MFPT} but save the analysis for the following section where we can compare this desorption protocol to desorption-induced resetting.

 \begin{figure*}[t!]
\centering
\includegraphics[width=16cm]{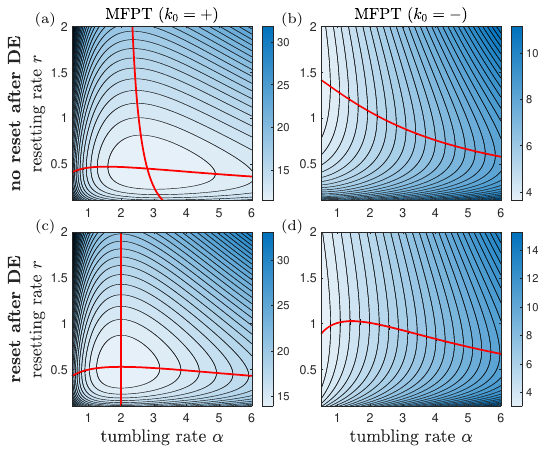}
\caption{Contour plots in the $(\alpha,r)$ plane of the absorption MFPT with non-resetting desorption (top row) and desorption-induced resetting to the initial state $(x_0,k_0)$ (bottom row) . In (a,c) $k_0=+$; in (b,d) $k_0=-$. Minimal curves are indicated in red. Other parameters are $x_0=1$, $v=2$, $\gamma_0=\overline{\gamma}_0=1$.}
\label{MFPT_contour}
\end{figure*}

\subsection{Reset after desorption}
As in Sect.\ III, the analysis simplifies if the particle immediately resets to $(x_0,k_0)$ after desorption, see scenario (i) in Fig. \ref{fig5}. Setting $(x^*,k^*)=(x_0,k_0)$ in the general renewal Eq.\ \eqref{srenf}, we now have
 \begin{align}
   \label{ren20}
  &\calF_{r,k_0}(x_0,t)=\pi_b\int_0^t\dd \tau  {\phi}(t-\tau) f_{r,k_0}(x_0,\tau) \\
&+\pi_d \int_0^t\dd\tau' \int_{\tau'}^t \dd\tau\,  {\calF}_{r,k_0}(x_0,t-\tau) \phi(\tau-\tau') f_{r,k_0}(x_0,\tau'). \nonumber
 \end{align}
Taking the Laplace transform of Eq.\ \eqref{ren20} yields
\begin{align}
	\label{wcalF}
	\widetilde{\calF}_{r,k_0}(x_0,s) = \frac{\pi_b \wphi(s)\f_{r,k_0}(x_0,s)}{1 - \pi_d \wphi(s)\f_{r,k_0}(x_0,s)},
\end{align}
which we use to determine the FPT moments,
\begin{align}
	\fT_{r,k_0}^{(n)}(x_0) =\Big( \!-\!\frac{d}{ds} \Big)^n \widetilde{\calF}_{r,k_0}(x_0,s) \Big|_{s=0}.
\end{align}
Assuming again that $\phi$ is not heavy-tailed, we substitute the relevant series expansions from Eq.\ \eqref{srs} into Eq.\ \eqref{wcalF} and Taylor expanding in powers of $s$ to find
\begin{align}
	\label{mfpt}
	\fT^{(1)}_{r,k_0}(x_0) = \calT_{r,k_0}(x_0) + \langle\tau\rangle + \Gamma\big(\calT_{r,k_0}(x_0) + \langle\tau\rangle\big).
\end{align}
That is, the mean FPT for absorption is the sum of the mean adsorption time from $x_0$, the mean absorption time following the final adsorption event, and the mean time accrued due to desorption events.

As before, collecting second-order terms in the Taylor expansion of Eq.\ \eqref{wcalF} yields the second-order FPT moment,
\begin{align}
	\label{mfpt2}
	\fT^{(2)}_{r,k_0}(x_0) = &(1+\Gamma)\big(\calT^{(2)}_{r,k_0}(x_0) + \langle\tau^2 \rangle +2\langle\tau\rangle \calT_{r,k_0}(x_0)\big)\nonumber\\
 &+ 2\Gamma(1+\Gamma)\big( \calT_{r,k_0}(x_0)+\langle\tau\rangle \big)^2.
\end{align}

In the absence of desorption we reproduce Eq.\ \eqref{vargam0}. It is critical to note, however, that while the absorption MFPTs under both desorption protocols depend on the waiting time density only with respect to its average, the choice of waiting time density is realized in higher-order moments. In particular, by allowing $\phi$ to be non-exponential, we can independently vary the mean and variance of the waiting time density.

  \begin{figure}
\centering
\includegraphics[width=8cm]{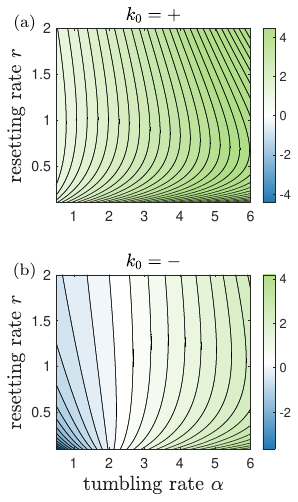} 
\caption{Contour plots in the $(\alpha,r)$ plane of the difference in absorption MFPTs between desorption protocols for (a) $k_0=+$ and (b) $k_0=-$; green indicates that the non-resetting protocol is faster on average; blue indicates that the desorption-induced resetting protocol is faster on average. Other parameters are $x_0=1$, $v=2$, and $\gamma_0=\overline{\gamma}_0=1$.}
\label{fig9}
\end{figure}

 \begin{figure}[t!]
\centering
\includegraphics[width=8cm]{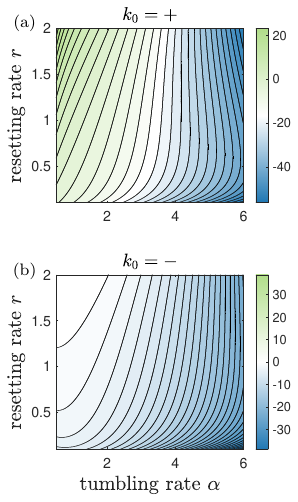} 
\caption{Contour plots in the $(\alpha,r)$ plane of the difference in absorption MFPTs between desorption protocols for (a) $k_0=+$ and (b) $k_0=-$ when desorption and bulk dynamics are made to match, that is, the desorption and bulk tumbling rates are equal ($\gamma_0=\alpha$) in the case of non-resetting desorption, and the desorption and bulk resetting rates are equal ($\gamma_0=r$) in the case of resetting-induced desorption; green indicates that the non-resetting protocol is faster on average; blue indicates that the desorption-induced resetting protocol is faster on average. Other parameters are $x_0=1$, $v=2$, and $\overline{\gamma}_0=1$.}
\label{fig10}
\end{figure}

\subsection{Results}

In Figs.\ \ref{fig:MFPT}(a,d) we plot the MFPT given by Eqs.\ \eqref{mfptnrad} and \eqref{mfpt}, respectively, as a function of the bulk resetting rate $r$. We observe the typical unimodal dependence of the MFPT on $r$. However, a number of other non-trivial results can be deduced. First, the optimal resetting rate $r^*$ is unaffected by $\Gamma$ when there is resetting after desorption. That is, bulk resetting is useful only to the extent that it can accelerate the time for the RTP to return to the boundary from $x_0$, whereas resetting also affects the desorbed state when it continues from the boundary. Second, while the MFPT with $k_0=-$ is less than that of $k_0=+$ in Figs.\ \ref{fig:MFPT}(a,d), the optimal bulk resetting rate (which we henceforth denote by $r^*$) is greater. This phenomenon reflects the fact that faster resetting rates can still be beneficial when $k_0=-$ since the RTP resets to the left-moving state. A similar story can be told about the influence of the tumbling rate on the absorption MFPT (see Figs.\ \ref{fig:MFPT}(b,e)) except unsurprisingly if $k_0=-$ the MFPT is minimized by eliminating tumbling altogether.

Perhaps the most interesting absorption MFPT behavior arises when the desorption protocol is tied to the bulk dynamics. In Fig.\ \ref{fig:MFPT}(c) we compare the absorption MFPT under no-reset desorption when the desorption and tumbling rates are the same. Likewise under the desorption-induced resetting protocol we compare the absorption MFPT when the desorption and bulk resetting rates are the same; see Fig.\ \ref{fig:MFPT}(f). In both scenarios it is immediate that these equivalences impose a trade-off wherein accelerating the adsorption time can also promote desorption over absorption. The resulting optimal resetting rate therefore lies somewhere in the range $[0,r^*]$ and, what is more, there exists a range of resetting rates exceeding $r^*$ for which the MFPT with $\gamma_0 = r$ remains less than the MFPT when the bulk and desorption resetting rates are independent (likewise for optimal tumbling with non-resetting desorption).

Resetting and tumbling kinetics aside, Figs.\ \ref{fig:NESS} and \ref{fig:MFPT} collectively highlight the strong influence of the initial and resetting orientation $k_0$ on the particle NESS and absorption MFPT. Figure \ref{MFPT_contour} summarizes these differences for the MFPT in the parameter space $(\alpha,r)$; for $k_0=\pm$ the MFPT depends unimodally on the bulk resetting rate $r$ for any fixed tumbling rate $\alpha$. (This optimal bulk resetting rate $r^*$ moreover depends non-monotonically on the tumbling rate, see pseudo-horizontal red curves.) However, the converse is only true of $k_0=+$ hence in this case there exists an optimal coordinate pair $(\alpha^*,r^*)\in \mathbb{R}^2_{>0}$ which minimizes the absorption MFPT; see Figs.\ \ref{MFPT_contour}(a,c). Note that $\alpha^*$ is invariant to $r$ only in the desorption-induced resetting protocol.

In Fig.\ \ref{fig9} we illustrate the difference in the contour plots of Fig.\ \ref{MFPT_contour} between desorption protocols. That is, the data in Fig.\ \ref{fig9}(a) is that of Fig.\ \ref{MFPT_contour}(c) subtracted from that of Fig.\ \ref{MFPT_contour}(a); likewise for Fig.\ \ref{fig9}(b) and Figs.\ \ref{MFPT_contour}(d) and \ref{MFPT_contour}(b). For an initially right-moving ($k_0=+$) RTP the absorption MFPT with the non-resetting desorption protocol is always faster than that of the desorption-induced resetting protocol; see Fig.\ \ref{fig9}(a). This result is further enhanced by higher tumbling rates which disproportionately facilitate boundary interactions for an RTP that does not reset following desorption. This behavior does not come as a surprise, however, since the bound time fractions for the non-absorbing boundary setting with $k_0=+$ and $v=2$ in Figs.\ \ref{fig:BTF}(c,d) lie exclusively above unity for all tumbling and resetting rates.

The case of an initially left-moving ($k_0=-$) RTP is markedly different; see Fig.\ \ref{fig9}(b). Here we see that the desorption-induced resetting protocol yields faster absorption MFPTs so long as the tumbling rate $\alpha$ is not too large. This phenomenon is explained by the fact that desorption induces subsequent movement back toward the boundary whereas the non-resetting desorption protocol causes subsequent movement away from the boundary. We also see this distinction reflected in the corresponding bound time fraction with $k_0=-$ and $v=2$; see Figs.\ \ref{fig:BTF}(c,d).

Figure \ref{fig10} similarly illustrates contour plots in the $(\alpha,r)$ plane of the difference in absorption MFPTs between desorption protocols except now we take the desorption rate of the non-resetting protocol to be equal to the bulk tumbling rate $\gamma_0=\alpha$ and likewise for resetting-induced absorption with $\gamma_0=r$. Comparing Figs.\ \ref{fig9} and \ref{fig10} for a given initial orientation $k_0$ reveals almost entirely opposite `preferred' desorption protocols for minimizing the absorption MFPT. What we observe is that too large of a tumbling rate makes absorption progressively less likely for the non-resetting desorption protocol. Similarly, too large of a resetting rate makes absorption unlikely for the desorption-induced resetting protocol.

\setcounter{equation}{0}
 
 \section{Discussion}
In this paper we developed a general renewal formalism to characterize the effects of stochastic resetting and a sticky boundary on the dynamics of an RTP on the half line. We first assumed the boundary at $x=0$ to be non-absorbing and the particle non-resetting. Upon colliding with the boundary (adsorption) the RTP adhered for a random waiting time before switching directions and re-entering the bulk (desorption). We then considered the RTP to instantaneously reset to its initial position and orientation while navigating the bulk domain. During desorption, the RTP could either re-enter the domain from $x=0$ as before or reset as in the bulk. Finally we determined the FPT statistics of the stochastically resetting RTP when allowing for absorption at the boundary. 

The technical development of this work lies in the utility of the renewal formalism to characterize resetting stochastic search processes among sticky boundaries in terms of successive resetting events and boundary collisions. That is, we analysed the probability density of the resetting RTP in terms of sequential excurions of an RTP without resetting and, similarly, the absorption time density in terms of sequential desorption events. The renewal approach is not only oftentimes simpler to work with, as evident in Sect. II, it also seamlessly accommodates more general waiting time models, e.g.\ non-Markovian models of resetting procedures, desorption, and absorption.

For the sake of illustration, we restricted the analysis of our model to Markovian resetting and boundary interactions. In doing so we observed in the RTP significant asymmetries with respect to the initial orientation . We found these asymmetries in the NESS (in the absence of absorption) to be accentuated by the proximity of $x_0$ to the boundary and by the bulk resetting rate $r$. Pronounced differences also occurred with regards the average bound time fraction: for $k_0=+$, the non-resetting desorption protocol always exhibits a higher bound time fraction, but for $k_0\!=\! -$ the behavior is much more sublte.

With either initial orientation we found that the absorption MFPT depends unimodally on the resetting rate and, with all other parameters the same, a smaller optimal resetting rate for $k_0=+$ than $k_0=-$ since resetting then induces movement away from the boundary. More surprisingly, we found interactions between the optimal bulk resetting rate $r^*$ and tumbling rate $\alpha$ to be nontrivial, with $r^*$ evolving either non-monotonically or unimodally with respect to $\alpha$ depending on the initial orientation and desorption protocol. When $k_0=+$ there is a strictly positive global optimal coordinate pair $(\alpha^*,r^*)$ which minimizes the absorption MFPT. These optima differ between desorption protocols. Finally, while the non-resetting desorption protocol always yields a faster MFPT when $k_0= +$ (and keeping other parameters fixed), the desorption-induced resetting protocol can produce faster MFPTs when $k_0=-$ so long as the tumbling rate is not too large. Thus, both bulk resetting and desorption resetting can serve to accelerate absorption at the boundary.

As a final investigation into the behavior of a resetting RTP with a sticky boundary, we considered (i) the non-resetting desorption protocol with the same desorption and bulk tumbling rates ($\gamma_0=\alpha$) and (ii) the desorption-induced resetting protocol with the same desorption and bulk resetting rates ($\gamma_0=r$). The motivation behind the first scenario comes from the fact that boundary behavior of an RTP in higher spatial dimensions is usually characterized by the continuously evolving tumbling direction that the particle exhibits in the bulk. In each case, minimizing the MFPT involves a trade-off between optimizing collision times with the boundary at rate $\alpha^*$ (likewise $r^*$) and promoting absorption over desorption with $\gamma=0$. Ultimately the optimal rates in these settings lie within the range $(0,\alpha^*)$ (likewise $(0,r^*)$). Finally, perhaps the most significant result of the work is that for any parameter set $(\alpha,r,k_0)$, the desorption protocol that minimizes the MFPT when desorption and bulk kinetics are made equal is almost always the opposite of when desorption and bulk dynamics are distinct. In short, desorption dynamics factor critically yet quite subtly into the story of a resetting RTP with a sticky boundary.

There are a number of possible extensions of the current work. First, we could include the effects of diffusion on an RTP in 1D along the lines of Refs. \cite{Malakar18,Bressloff25}. However, this would require modifying the derivation of the renewal equation without desorption-induced resetting, since the boundary is totally adsorbing with respect to both the left-moving and right-moving states following each arrival event. The same issue would also apply to other models of an active particle in a semi-infinite channel such as an active Brownian particle (ABP) and an active Ornstein-Uhlenbeck particle (AOUP) \cite{Ramaswamy10,Vicsek12, Roman12,Bechinger16}. Second, we could reinterpret an RTP as an active version of a random searcher such as a foraging animal in which the boundary is a partially accessible target \cite{Bressloff25b}. In this case, desorption-induced resetting would correspond to the searcher returning to its home base each time it fails to access the target (no absorption). From the perspective of animal foraging, a major simplification of our current model is to assume that the searcher is memoryless. That is, the searcher does not modify its bulk dynamics and whenever it returns to the target it interacts in the same way as previous visits. One biological motivation for this, at least in the case of single-cell organisms, is that maintaining a memory of previous searches costs energy. Within the general theory of active search processes, a challenging problem is to extend the model to multiple targets in higher spatial dimensions. There are several difficulties that need to be addressed. First, analysing the bulk dynamics of an active particle in higher dimensions is non-trivial.  Second, as previously shown for diffusive search processes in $\R^d$, $d >1$ \cite{Bressloff25a,Bressloff25b}, the renewal equations involve spatial integrals with respect to points on the $d-1$-dimensional target surfaces. Third, finding the most efficient search protocol will depend on how the targets are distributed across the search domain, the location of the home base, and the desorption protocol at each target. Fourth, the relevant timescale may not be the single-target search time but rather the time taken to find every target. These `cover time' statistics have been characterized for a memoryless RTP in the frequent resetting limit \cite{Linn25} but more general results remain elusive. Finally, one could study the model within the context of resource accumulation where resources are sequentially generated at the target sites; in this case a recently developed queuing theoretic framework would be well-suited to the analysis \cite{JGB25}.

\section{Data availability}
No data were created or analysed in this study.

\section{Acknowledgements} SL was supported by the U.S.\ National Science Foundation grant DMS-2503350.

\setcounter{equation}{0}
\renewcommand{\theequation}{A.\arabic{equation}}
\section*{Appendix A: Derivation of the renewal equations}

In this Appendix we derive the renewal equations for the propagator and FPT density in the case of a partially absorbing sticky boundary. For the sake of generality, we assume that there is both bulk resetting to the initial state $(x_0,k_0)$ and desorption-induced resetting to a state $(x^*, k^*)$. In the main text we take $(x^*,k^*)=(x_0,k_0)$. The case of no desorption-induced resetting is recovered by setting $(x^*,k^*)=(0,+)$. Let $X(t)\in [0,\infty)$ denote the position of the RTP at time $t$ and let $\sigma(t)$ denote the corresponding velocity state. The initial conditions are $X(0)=x_0$ and $\sigma(0)=k_0$. Given the FPT for absorption $\calT$, the propagator can be expressed as 
\begin{equation}
\rho_{r,k|k_0}(x,t|x_0)=\E[\delta(x-X(t))\delta_{k,\sigma(t)}{\bf 1}_{\calT>t}]
\end{equation}
where ${\bf 1}$ is the indicator function and $\E[\cdot]$ denotes a double expectation with respect to tumbling, resetting and the desorption/absorption process. Similarly let $\overline X(t)$ denote the  position of an RTP in the case of a totally absorbing (non-sticky) boundary with corresponding FPT $\overline \calT$ such that
\begin{equation}
p_{r,k|k_0}(x,t|x_0)=\E[\delta(x-\overline X(t))\delta_{k,\sigma(t)}{\bf 1}_{\overline \calT>t}].
\end{equation}

\subsection*{A.1 Renewal equation for the propagator}

In order to derive a renewal equation relating $\rho_{r,k|k_0}$ to $p_{r,k|k_0}$,  we condition the total expectation with respect to the first arrival time $\tau_f$ of the RTP at the boundary $x=0$. Frst consider the decomposition
\begin{align}
\label{s1}
 \rho_{k|k_0}(x,t|x_0)&= \E[\delta(x-X(t))\delta_{k,\sigma(t)}{\bf 1}_{\calT>t} |\tau_f > t]    \\
&\qquad  +\,  \E[\delta(x-X(t))\delta_{k,\sigma(t)}{\bf 1}_{\calT>t}|\tau_f\leq t ].\nonumber 
\end{align}
If $\tau_f>t$ then $\calT>t$ is guaranteed and $X(t)$ has the same statistics as $\overline X(t)$ with $\overline \calT >t$. Hence, 

\begin{eqnarray}
\label{s2}
 &&\E[\delta(x-X(t))\delta_{k,\sigma(t)}{\bf 1}_{\calT>t} |\tau_f > t] \\
 & & \qquad=\E[\delta(x-\overline X(t))\delta_{k,\sigma(t)}{\bf 1}_{\overline \calT>t}]  
   =p_{r,k|k_0}(x,t|x_0). \nonumber 
\end{eqnarray}

On the other hand, if $\tau_f< t$ then the RTP binds to the wall at time $\tau_f$. After a random waiting time $\tau_1$ with probability density $\phi(\tau)$, the particle either desorbs and resets to $(x^*,k^*)$ with probability $\pi_d$ or is permanently absorbed with probability $\pi_b$. We thus have the renewal relation
 \begin{equation}
 \label{Xads_des}
 X(t)=\left \{ \begin{array}{ll} 0 &\mbox{ for } \tau_f+\tau_1 <t\\
\widehat X(t-\tau_f-\tau) &  \mbox{ for } \tau_f+\tau_1 >t \mbox{ with Prob. } \pi_d\\
\emptyset &  \mbox{ for } \tau_f+\tau_1 >t \mbox{ with Prob. } \pi_a \end{array}\right .
\end{equation}
Here $\emptyset$ represents the permanently absorbed state whereas the stochastic process $\widehat X(t)$ has the same statistics as $X(t)$ under the initial condition $\widehat X(0)=x^*$. Moreover, the arrival distribution for $\tau_f$ is independent of $X(t)$ and is given by the FPT density $f_{r,k_0}(x_0,t')$ of $\overline \calT$. Finally, if the RTP desorbs then $\sigma(t)=\widehat{\sigma}(t-\tau_f-\tau_1)$ with $\widehat \sigma (t)$ having the same statistics as $\sigma(t)$ under the initial condition $\sigma(0)=k^*$. 

Hence, conditioning with respect to $\tau_f=\tau'$ and $\tau_1=\tau''$ we have

\begin{widetext}
\begin{eqnarray}
  \E[\delta(x-X(t))\delta_{k,\sigma(t)}{\bf 1}_{\calT>t} |\tau_f \leq t] 
  =\int_0^{\infty} \dd t''\,\phi(t'')  \int_0^{t}\dd t' \, \E[\delta(x-X(t))\delta_{k,\sigma(t)}{\bf 1}_{\calT>t} |\tau_1=t',\sigma_1=t'']f_0(x_0,t') 
  \label{s3}
 \end{eqnarray}
 with
\begin{eqnarray}
&&\E[\delta(x-X(t)\delta_{k,\sigma(t)}{\bf 1}_{\calT>t}| \tau_1=t',\sigma=t'']\nonumber \\
 &&= \Theta(\tau'+\tau''-t)\delta(x)+\pi_d \E[\delta (x-\widehat{X}(t-\tau'-t\tau'))\delta_{k,\widehat{\sigma}(t-\tau'-\tau')}{\bf 1}_{\widehat \calT>t}] \Theta(t-\tau'-\tau'')\nonumber \\
 && =\Theta(\tau'+\tau'-t)\delta(x)+\pi_d \rho_{r,k|k^*}(x,t-\tau'-\tau''|x^*) \Theta(t-\tau'-\tau'') ,
 \label{s4}
 \end{eqnarray}
 where $\Theta(x)$ is a Heaviside function. Substituting Eqs.\ \eqref{s2}--\eqref{s4} into \eqref{s1} yields the renewal equation
 \begin{align}
\rho_{r,k|k_0}(x,t|x_0)&= p_{r,k|k_0}(x,t|x_0) \\
&\qquad +\int_{0}^t \dd\tau' \bigg[ \pi_d \int_0^{t-\tau'}\dd\tau''  \rho_{k|k^*}(x,t-\tau'-\tau''|x^* )\phi(\tau'') \dd\tau''+\delta(x)\int_{t-\tau'}^{\infty} \phi(\tau'')\dd\tau''\bigg ]  f_{r,k_0}(x_0,\tau').  \nonumber
 \end{align}
 Finally, if we exclude the contribution at the boundary $x=0$ due to stickiness and perform the change of variables $\tau=\tau''+\tau'$, we obtain the renewal equation used in the main text,
 \begin{align}
\label{srenp}
  \rho_{r,k|k_0}(x,t|x_0)=p_{r,k|k_0}(x,t|x_0)
  +\pi_d\int_0^t\dd\tau' \int_{\tau'}^t \dd\tau\,  \rho_{r,k|k^*}(x,t-\tau|x^*)  \phi(\tau-\tau') f_{r,k_0}(x_0,\tau').  \end{align}
  \end{widetext}
 
\subsubsection*{Renewal equation for the FPT density} 
The FPT density for absorption has the representation
\begin{equation}
\calF_{r,k_0}(x_0,t)= \E[\delta(t-\calT)].
\end{equation}
Proceeding along analogous lines the derivation of the propagator, we require $\calT > \tau_f$ since absorption can occur only if the particle is attached to the boundary at $x=0$. Taking $\tau_f <t$ and conditioning with respect to both $\tau_f$ and the waiting time $\tau_1$ then gives
\begin{align}
 \calF(x_0,t)&=\int_0^{\infty}\dd\tau'' \, \phi(\tau'')\int_0^{t}\dd\tau'  \\
 &\quad \times\, \E[\delta(\calT-t)| \tau_f=\tau',\tau_1=\tau']  f_0(x_0,\tau').\nonumber
\end{align}

Either absorption occurs with probability $\pi_b$ at the time $\calT=\tau_f+\tau_1$ or with probability $\pi_d=1-\pi_b$ it desorbs and reenters the bulk domain so that $\tau_f+\tau_1<\calT$. Hence we have the renewal relation
\begin{equation}
 \label{Xhatreset2}
 \calT=\left \{ \begin{array}{ll} \tau_f+\tau_1 &\mbox{ with probability } \pi_b\\
\widehat \calT+\tau_f+\tau_1 & \mbox{ with probability } \pi_d \end{array}.\right .
\end{equation}
where $\widehat \calT$ is the FPT of $\widehat{X}(t)$. It follows that
\begin{widetext}
\begin{eqnarray}
 \calF_{r,k_0}(x_0,t)&=&\int_0^{\infty}\dd\tau'' \, \phi(\tau'')\bigg [\int_0^{t}\dd\tau' \,  \bigg([\pi_b \delta(t-\tau'-\tau'')+\pi_d \E[\delta ( \widehat \calT+\tau'+\tau''-t)]\bigg ) f_{r,k_0}(x_0,\tau')\\
 & =&\pi_b\int_0^t\dd\tau'  {\phi}(t-\tau'')  f_{r,k_0}(x_0,\tau')+\pi_d \int_0^t\dd\tau' \int_{0}^{t-t'} \dd\tau\, \calF_{r,k^*}(x^*,t-\tau'-\tau'') \phi(t'')  f_{r,k_0}(x_0,\tau').\nonumber
 \label{Fren_ads_des}
\end{eqnarray}
Finally performing the change of variables $\tau=\tau'+\tau''$ yields the FPT renewal equation used in the main text,
 \begin{align}
  \calF_{r,k_0}(x_0,t)=\pi_b\int_0^t\dd \tau ' {\phi}(t-\tau') f_{r,k_0}(x_0,\tau')    
+\pi_d \int_0^t\dd\tau' \int_{\tau'}^t \dd\tau\,  {\calF}_{r,k^*}(x^*,t-\tau) \phi(\tau-\tau') f_{r,k_0}(x_0,\tau'). \label{srenf}
 \end{align}
\end{widetext}

\setcounter{equation}{0}
\renewcommand{\theequation}{B.\arabic{equation}}
\section*{Appendix B: Calculation of $\p_{r,k|k_0} $}

Laplace transforming Eq.\ \eqref{CKHH2} with $(y_0,j_0)=(x_0,k_0)$ yields
\begin{align}
&  vk\frac{\partial \p_{r,k|k_0} }{\partial x}+(\alpha +r+s)\p_{r,k|k_0}  
  -\alpha \p_{r,-k|k_0}  \nonumber \\
  &\qquad =\delta_{k,k_0}\bigg (1+r\S_{r,k_0}(x_0,s)\bigg )\delta(x-x_0),
  \label{rLTP}
\end{align} 
with the boundary condition 
 $\p_{r,+|k_0}(0,s|x_0)=0$.
Setting $k=-k_0$ in Eq.\ \eqref{rLTP} expresses $\p_{r,k_0|k_0} $ in terms of $\p_{r,-k_0|k_0} $ according to
 \begin{align}
 \label{ord1}
\p_{r,k_0|k_0} = \frac{v}{\alpha} \bigg [ -k_0\frac{\partial \p_{r,-k_0|k_0} }{\partial x}+\lambda(s+r)\p_{r,-k_0|k_0} \bigg ] 
  \end{align}
 where $\lambda(s)=(\alpha +s)/v$. In order to solve for $\p_{r,k_0|k_0} $, we multiply both sides of Eq.\ \eqref{ord1} by the operator $k_0v\partial_x +(\alpha+r+s)$ and use Eq.\ \eqref{rLTP} with $k=k_0$. After some algebra we find that
 \begin{align}
&  \frac{\partial^2 \p_{r,-k_0|k_0} }{\partial x^2}-\Lambda(r+s)^2\p_{r,-k_0|k_0}  
  \nonumber \\
  &\qquad =-\frac{\alpha}{v^2}\bigg (1+r\S_{r,k_0}(x_0,s)\bigg )\delta(x-x_0)
  \label{ord2}
\end{align} 
 with
 \begin{equation}
 \Lambda(s)=\frac{1}{v}\sqrt{s(2\alpha+s)}.
 \end{equation}
Equation \eqref{ord2} has the general solution
\begin{align}
 \label{pLT}
\p_{r,-k_0|k_0}(x,s|x_0)&=\frac{1}{v}\left (1+r {\S}_{r,k_0}(x_0,s)\right )G(x,r+s|x_0)\nonumber \\
&\quad +\delta_{k_0,+}A(x_0,s)\e^{-\Lambda(r+s)x}
\end{align}
where
\begin{equation}
G(x,s|x_0)=\frac{\alpha}{2v\Lambda(s)}\bigg [\e^{-\Lambda(s)|x-x_0|} -\e^{-\Lambda(s)(x+x_0)}\bigg ]
\end{equation}
with $A(x_0,s)$ and the survival probability ${\S}_{r,k_0}(x_0,s)$ determined self-consistently using the Laplace transform of Eq.\ \eqref{surv},
\begin{equation}
\label{Sr}
{\S}_{r,k_0}(x_0,s)=\frac{1-v\p_{r,-|k_0}(0,s|x_0)}{s}.
\end{equation}

First consider the case $k_0=+$ for which
\begin{equation} \label{fr}
  \f_{r,+}(x_0,s)\equiv 1-s{\S}_{r,+}(x_0,s)=vA(x_0,s).
\end{equation}
In addition, taking the limit $x\rightarrow 0$ in Eq.\ \eqref{rLTP} for $k_0=+$ implies that
\begin{align}
0&=  -\frac{\partial \p_{r,-|+}(0,s|x_0) }{\partial x}+\lambda(r+s)\p_{r,-|+}(0,s|x_0)\nonumber \\  
  &=-\frac{\alpha}{v^2}(1+rS_{r,+}(x_0,s))\e^{-\Lambda(r+s)x_0}\nonumber \\
  &\qquad +\left (\Lambda(r+s)+\lambda(r+s) \right )A(x_0,s).
\end{align} 
 Hence,
\begin{align}
A(x_0,s) =\frac{\alpha}{v^2}\frac{(r+s)\e^{-\Lambda(r+s)x_0}}{s[\Lambda(r+s)+\lambda(r+s)]+(\alpha r/v)\e^{-\Lambda(r+s)x_0}}.
\end{align}
It also follows that
\begin{align}
\lim_{s\rightarrow 0} A(x_0,s)=v^{-1}
\end{align}
and from Eq.\ \eqref{fr}

\begin{align}
\label{Tplus}
T_{r,+} (x_0)&=-v\left . \frac{d}{ds} A(x_0,s) \right |_{s=0}\nonumber \\
&=\frac{1}{r}\left (\frac{v}{\alpha}\bigg [\Lambda(r)+\lambda(r)\bigg ]\e^{\Lambda(r)x_0}-1\right )\\
&=\frac{1}{r}\left (\frac{\sqrt{r(2\alpha+r)}+\alpha+r}{\alpha} \e^{\sqrt{r(2\alpha+r)}x_0/v}-1\right ).\nonumber
\end{align}
Turning to the case $k_0=-$, Eq.\ \eqref{pLT} becomes
\begin{align} \label{pLTpm}
\p_{r,+|-}(x,s|x_0)&=\frac{1}{v}\left (1+r {\S}_{r,-}(x_0,s)\right )G(x,r+s|x_0).
\end{align}
 Moreover from Eq.\ \eqref{ord1} we have
 \begin{align}
 \label{ord3}
&\p_{r,-|-} (x,s|x_0) = \frac{1}{\alpha}  \left (1+r {\S}_{r,-}(x_0,s)\right ) \\
&\qquad \times\,\bigg [ \frac{\partial G(x,r+s|x_0) }{\partial x}+\lambda(s+r)G(x,r+s|x_0) \bigg ] \nonumber.
  \end{align}
Equation \eqref{Sr} leads to the implicit equation
\begin{align}
1-s{\S}_{r,-}(x_0,s)&=\frac{v}{\alpha}  \left (1+r {\S}_{r,-}(x_0,s)\right ) \frac{\partial G(0,r+s|x_0) }{\partial x}\nonumber \\
&=  \left (1+r {\S}_{r,-}(x_0,s)\right ) \e^{-\Lambda(r+s)x_0}.
\end{align}
Rearranging gives
\begin{align}
{\S}_{r,-}(x_0,s)=\frac{1- \e^{-\Lambda(r+s)x_0}}{ s+r \e^{-\Lambda(r+s)x_0} }. \label{Srmx0s}
\end{align}
Finally taking $s\to 0$ in Eq.\ \eqref{Srmx0s} we determine
\begin{align}
\label{Tminus}
T_{r,-}(x_0)=\frac{1}{r} \left [ \e^{\Lambda(r)x_0}-1\right ].
\end{align}


\begin{thebibliography}{100}
\bibitem{Ramaswamy10} S. Ramaswamy. The mechanics and statistics
of active matter. {Annu. Rev. Condens. Matter Phys.} {\bf 1}, 323-345 (2010)

\bibitem{Vicsek12} T. Vicsek and A.Zafeiris. Collective motion. {Phys. Rep.}
{\bf 517}, 71-140 (2012).

\bibitem{Roman12} R. Romanczuk, M. Bar, W. Ebeling, B. Lindner and
L. Schimansky-Geier. Active Brownian particles:
From individual to collective stochastic dynamics. {Eur. Phys. J. Special Topics }{\bf 202}, 1-162 (2012)

\bibitem{Bechinger16} G. Bechinger, R. Di Leonardo, H. Lowen, C. Reichhardt, G. Volpe  and G. Volpe. Active particles in complex and crowded environments. {Rev. Mod. Phys.} {\bf 88}, 045006 (2016)

\bibitem{Angelani17} L. Angelani. Confined run-and-
 swimmers in one dimension. {J. Phys. A} {\bf 50} 325601 (2017)

\bibitem{Angelani23} L. Angelani. One-dimensional run-and-tumble motions
with generic boundary conditions. {\em J. Phys. A} {\bf 56} 455003 (2023)

 \bibitem{Malakar18} K. Malakar, V. Jemseena, A. Kundu, V Kumar, 
S. Sabhapandit, S. N. Majumdar, S. Redner and A. Dhar. Steady state, relaxation and first-passage properties of a run-and-tumble particle in one-dimension, {J. Stat.
Mech.} 043215 (2018).

\bibitem{Bressloff25} P. C. Bressloff. Run-and-tumble particle with diffusion: boundary local times and the zero-diffusion limit. {J. Stat.Mech.} {\bf 113201} (2025).

\bibitem{Bressloff23} P. C. Bressloff. Encounter-based model of a run-and-tumble particle II: absorption at sticky boundaries. J. Stat. Mech. {\bf 043208}  (2023).

\bibitem{Bressloff25d} P. C. Bressloff. Stochastic calculus of run-and-tumble motion: an applied perspective. Proc. Roy Soc. A 481 20240815 (2025)

 \bibitem{Evans18} M. R. Evans and S. N. Majumdar. Run and tumble particle under resetting: a renewal approach. {J. Phys. A: Math. Theor.} {\bf 51} 475003 (2018).
  
\bibitem{Bressloff20} P. C. Bressloff, Occupation time of a run-and-tumble particle with resetting.  {Phys. Rev. E} {\bf 102} 042135  (2020).

\bibitem{Santra20a} I. Santra, U. Basuand S. Sabhapandit, Run-and-tumble particles in two dimensions under stochastic resetting conditions {J. Stat. Mech.} {\bf 113206} (2020).

\bibitem{Bressloff25c}  P. C. Bressloff. Encounter-based model of a run-and-tumble particle with stochastic resetting. J. Phys. A 58 125002 (2025)

\bibitem{Dhar19} A, Dhar, A. Kundu, S. N. Majumdar, S. Sabhapandit  and G. Schehr. Run-and-tumble particle in one-dimensional confining potentials: Steady-state, relaxation,
and first-passage properties. Phys. Rev. E {\bf 99}, 032132 (2019)

\bibitem{Bressloff25a} P. C. Bressloff. Diffusion-mediated adsorption versus absorption at partially reactive targets: a renewal approach. J. Phys. A 58 245003 (2025)

\bibitem{Bressloff25b} P. C. Bressloff. Random search with stochastic resetting: when finding the target is not enough. Phys. Rev. 111 054127 (2025)

\bibitem{Grebenkov23} D. S. Grebenkov, Diffusion-controlled reactions with non-Markovian
binding/unbinding kinetics {J. Chem. Phys.} {\bf 158} 214111 (2023).

\bibitem{Scher23} Y.Scher, S. Reuveni and D. S. Grebenkov
Escape of a sticky particle. Phys. Rev. Res.
{\bf 5} 043196 (2023)

\bibitem{Linn25} S. Linn and S. D. Lawley. Cover times with stochastic resetting. {Chaos} {\bf 35}(4), 043124 (2025).
  
\bibitem{JGB25} J Giral-Barajas and P. C. Bressloff. Stochastic 1D search-and-capture as a \emph{G/M/c} queueing model.  {\em J. Phys. A: Math. Theor.} {\bf 58} 355001 (2025).

\end{thebibliography}
\end{document}